\documentclass[reprint,superscriptaddress,amsmath,amssymb,aps]{revtex4-2}
\usepackage{graphicx}
\usepackage{bm}
\usepackage{physics}
\usepackage{hyperref}
\usepackage{amsfonts}
\usepackage{xcolor}
\usepackage{float}
\usepackage{comment}
\usepackage{enumitem}
\begin{document}

\title{Programmable pixel-mode linear interferometers using multi-plane light conversion}

\author{Mushkan Sureka}
\thanks{These authors contributed equally to this work. See Author Contributions section for further details.}
\affiliation{Department of Electrical and Computer Engineering, University of Maryland, College Park MD}
\author{Itay Ozer}
\thanks{These authors contributed equally to this work. See Author Contributions section for further details.}
\affiliation{Department of Electrical and Computer Engineering, University of Maryland, College Park MD}
\author{Wenhua He}
\affiliation{Department of Electrical and Computer Engineering, University of Maryland, College Park MD}
\author{Michael R. Grace}
\thanks{Now at RTX BBN, Cambridge MA.}
\affiliation{Wyant College of Optical Sciences, University of Arizona, Tucson AZ}
\author{Chaohan Cui}
\affiliation{Department of Electrical and Computer Engineering, University of Maryland, College Park MD}
\author{Saikat Guha}
\affiliation{Department of Electrical and Computer Engineering, University of Maryland, College Park MD}
\affiliation{Wyant College of Optical Sciences, University of Arizona, Tucson AZ}

\begin{abstract}
Programmable linear optical interferometers are a core primitive in optical signal processing, quantum information processing, and photonic computing. Existing photonic-integrated implementations realize arbitrary $M$-mode unitaries using Mach--Zehnder-interferometer meshes whose footprint and accumulated loss scale with $O(M^2)$ optical components. Here we analyze and experimentally demonstrate a programmable architecture for implementing linear optical transformations directly on spatially tiled free-space {\em pixel modes} using multi-plane light conversion (MPLC). In this architecture, $M$ spatial modes arranged on a transverse lattice undergo a unitary transformation and are mapped to $M$ output modes of identical geometry through a sequence of programmable phase masks separated by free-space propagation segments. Numerical simulations show that arbitrary $M$-mode unitaries can be compiled to a desired high fidelity using a number of phase planes that scales approximately linearly with $M$. Using a spatial-light-modulator-based MPLC, we experimentally demonstrate programmable interferometers acting on up to $16$ spatial pixel modes, including tunable beamsplitters, Hadamard unitaries, spatial permutations, and partial unitaries on select subsets of modes. These results establish MPLC-based pixel-mode interferometers as a promising architecture for programmable linear optics with applications in classical and quantum optical interconnects, photonic switching, and quantum information processing.
\end{abstract}

\maketitle

\section{Introduction}

Linear optical interferometers capable of implementing programmable unitary transformations across multiple optical modes are a fundamental primitive for a wide range of photonic technologies. In classical optical systems, such transformations enable spatial multiplexing, optical switching, coherent signal processing, and photonic computing architectures, including those used for optical neural networks and machine-learning accelerators. In quantum photonics, programmable interferometers underpin applications such as boson sampling and Gaussian boson sampling, photonic quantum simulation, linear-optical quantum computing, generation of multimode continuous-variable entanglement, and joint-detection receivers for quantum-limited classical communication.

A universal $M$-mode linear optical interferometer implements a transformation
\begin{equation}
{\hat {\boldsymbol{a}}}_{\rm out} = U {\hat {\boldsymbol{a}}}_{\rm in},
\end{equation}
where ${\hat {\boldsymbol{a}}}_{\rm in}$ and ${\hat {\boldsymbol{a}}}_{\rm out}$ denote vectors of annihilation operators describing $M$ optical modes at the input and the output respectively, and $U$ is an $M\times M$ unitary matrix. Several universal architectures have been developed to realize such transformations. In particular, the decompositions introduced by Reck \textit{et al.}~\cite{reck1994} and later optimized by Clements \textit{et al.}~\cite{clements2016} show that arbitrary $M$-mode unitaries can be implemented using meshes of tunable beamsplitters and phase shifters arranged as Mach--Zehnder interferometers (MZIs). These architectures are widely implemented in photonic integrated circuits (PICs).

Although integrated interferometers provide excellent stability and scalability~\cite{Dong2023,Bogaerts2020} and have enabled demonstrations of large programmable photonic circuits~\cite{Wang2019,Luan2026}, their physical footprint and optical losses grow with the number of interferometric elements required to realize a general unitary transformation. Universal PIC interferometers require $O(M^2)$ tunable beamsplitters arranged in an $M\times M$ mesh~\cite{clements2016}. Each beamsplitter is implemented as an MZI containing directional couplers and phase shifters that must be tuned either thermally or electro-optically. As the circuit depth increases, accumulated propagation loss and fabrication tolerances become increasingly significant. In addition, the spatial locations and modal profiles of the optical modes are constrained by the waveguide geometry, and coupling losses often arise when interfacing free-space optical fields with waveguide modes due to mode mismatch.

An alternative approach to implementing complex spatial transformations on multimode optical fields is {\em multi-plane light conversion} (MPLC). MPLC systems consist of a sequence of $K$ phase masks separated by free-space propagation segments. By appropriately designing the phase profiles, MPLC can implement arbitrary unitary transformations across sets of orthogonal spatial modes. The concept was first introduced by Morizur \textit{et al.}~\cite{morizur2010}, who demonstrated programmable spatial mode manipulation using cascaded phase plates. Subsequent work by Labroille \textit{et al.}~\cite{labroille2014} and Fontaine \textit{et al.}~\cite{fontaine2019} established MPLC as a practical platform for high-dimensional spatial mode multiplexing and demultiplexing, enabling low-loss transformations among large sets of orthogonal spatial modes such as Laguerre--Gaussian and Hermite--Gaussian bases. 

More recently, MPLC systems have enabled demonstrations of very high mode-count spatial processors capable of sorting or multiplexing hundreds of spatial modes~\cite{labroille2014,fontaine2019,carpenter2015,dahl2020high}. These devices have been explored in a range of applications including spatial-division multiplexed optical communications, structured-light processing~\cite{fontaine2019}, quantum opto-mechanical systems~\cite{Choi2025}, and quantum-limited super-resolution imaging~\cite{Deshler2025,tan2023quantum}. In most of these demonstrations, however, MPLC devices implement fixed spatial transformations between orthogonal modal bases.

In this work we explore a different perspective: using MPLC as a {\em programmable interferometric processor} acting on spatially localized modes. Specifically, we consider optical modes arranged as spatially separated {\em pixel modes} on a two-dimensional lattice in the transverse plane. We develop and experimentally demonstrate a programmable architecture that realizes arbitrary linear-optical transformations across such pixel-mode arrays using a sequence of programmable phase masks in an MPLC system. Rather than decomposing the target unitary into a sequence of two-mode interferometers, the MPLC is used in its native diffractive form to implement the transformation directly.

This architecture offers several advantages relative to conventional waveguide-mesh interferometers. First, the implemented unitary is fully programmable through the phase masks of the MPLC system without requiring modifications to the physical optical hardware. Second, the spatial locations and mode shapes of the optical modes can be incorporated directly into the compilation process, enabling natural interfacing with free-space emitters such as atomic arrays~\cite{manetsch2025tweezer,holman2026trapping}, quantum dots, or VCSEL arrays. Third, the physical footprint of the interferometer is governed primarily by diffraction physics rather than waveguide geometry, potentially enabling compact implementations for large mode counts~\cite{wang2025ultracompact}.

We analyze this architecture through numerical simulations and experimental demonstrations. Using wavefront-matching compilation techniques, we evaluate the performance of MPLC implementations of several application-inspired unitaries, including beamsplitters, Hadamard transforms, spatial permutations, and linear-optical circuits used in boosted Bell-state measurements. A key result of our analysis is that the number of MPLC phase planes required to achieve a target fidelity scales approximately linearly with the number of modes, $K \propto M$. In Morizur \textit{et al.}'s original paper~\cite{morizur2010}, MPLC's universality---in performing an arbitrary unitary $U$ on any set of $M$ co-propagating orthonormal modes---was proved by first showing that a general beamsplitter can be performed on any pair of modes (while performing identity on the rest) using $O(1)$ planes, followed by arguing that $O(M^2)$ iterations of the above, using Clements {\em et al.}'s decomposition of an $M$-mode unitary into $O(M^2)$ two-mode beamsplitters~\cite{clements2016}, can help the MPLC realize any $M$-mode unitary. This argument works, but is highly superfluous as it uses $K = O(M^2)$ planes. Our finding of $K = O(M)$ suggests that the MPLC's native diffractive compilation of an $M$-mode unitary is far more powerful.

Experimentally, we implement programmable interferometers acting on up to sixteen spatial pixel modes using a spatial-light-modulator-based MPLC platform together with a reconfigurable coherent scene generator. We demonstrate tunable-transmissivity beamsplitters, Hadamard unitaries, spatial permutations, partial unitaries acting on selected subsets of modes.

A number of studies have investigated pixel-spatial-mode-based photonic processors for specific low-mode-count demonstrations, including Haar-random unitaries~\cite{lib2022pixel_mode_unitary}, cluster-state preparation~\cite{lib2024pixel_modes_q_computation}, and variational two-mode beam splitters~\cite{martinez2024pixel_modes}. Several novel platforms have been demonstrated acting on either pixel spatial modes or co-propagating orthogonal spatial modes, including phase-only spatial light modulators coupled with multimode fiber acting as complex scattering media~\cite{goel2024inversedesign_SLM_MMF}, metasurfaces~\cite{yousef2025metasurfaceHOM}, and spatial light modulators interleaved with free-space propagation~\cite{brandt2020HD_spatial_Qgate,wang2025ultracompact,Wang2024Ultrahigh-fidelityNetworks}.

The remainder of the paper is organized as follows. In Section~\ref{sec:architecture} we describe the pixel-mode MPLC architecture and the unitary compilation approach. Section~\ref{sec:simulations} presents numerical performance evaluations and scaling analysis. Section~\ref{sec:experiments} describes the experimental implementation and demonstrations. Finally, Section~\ref{sec:conclusions} discusses practical considerations and future directions for programmable spatial interferometers based on MPLC.

\section{Architecture}~\label{sec:architecture}

We consider $M$ spatially localized optical modes arranged on a regular square lattice in the transverse $(x,y)$ plane at the input plane $z=0$. Each input {\em pixel mode} is described by a normalized spatial field-shape $\phi_m(x,y) \equiv \phi_0(x-x_m,y-y_m)$ centered at a lattice site $(x_m,y_m)$, $1 \le m \le M$, where the pixel-mode's shape $\phi_0(x,y)$ is taken to be either a Gaussian or a hard-circular function, of mode-field diameter (MFD) $\sigma$ (mode-field standard-deviation in the case of Gaussian). The transverse center-to-center spacing between adjacent modes along the lattice directions is denoted by $b$. The dimensionless parameter $b/\sigma$ therefore characterizes the spatial separation between modes in units of the MFD. The goal of the device is to implement a linear optical transformation between these $M$ modes as input, and a matching set of modes $\psi_m(x',y') \equiv \phi_0(x'-x_m,y'-y_m)$ at the output plane of an MPLC (where $(x',y')$ are the transverse coordinates in that output plane), such that $\boldsymbol{\hat a}_{\rm out} = U \boldsymbol{\hat a}_{\rm in}$, where $U$ is an $M \times M$ unitary matrix describing the desired interferometer across the $M$ pixel modes indexed $1 \le m \le M$, with $\boldsymbol{\hat a}_{\rm in}$ and $\boldsymbol{\hat a}_{\rm out}$ denoting $M$-length annihilation-operator column vectors for the input and output pixel-mode sets respectively. The unitary transformation is implemented using a multi-plane light conversion system consisting of $K$ programmable phase masks separated by free-space propagation segments of length $w$ along the propagation direction, $z \in [0, L]$. Each phase mask consists of a $P \times P$ grid of phase-only pixels with physical size $\Delta \times \Delta$. The length of each transverse dimension of the modulated portion of each plane is therefore $H = P \Delta$. The optical field propagates between successive planes according to Rayleigh--Sommerfeld diffraction.
\begin{figure}
    \centering
    \includegraphics[width=\columnwidth]{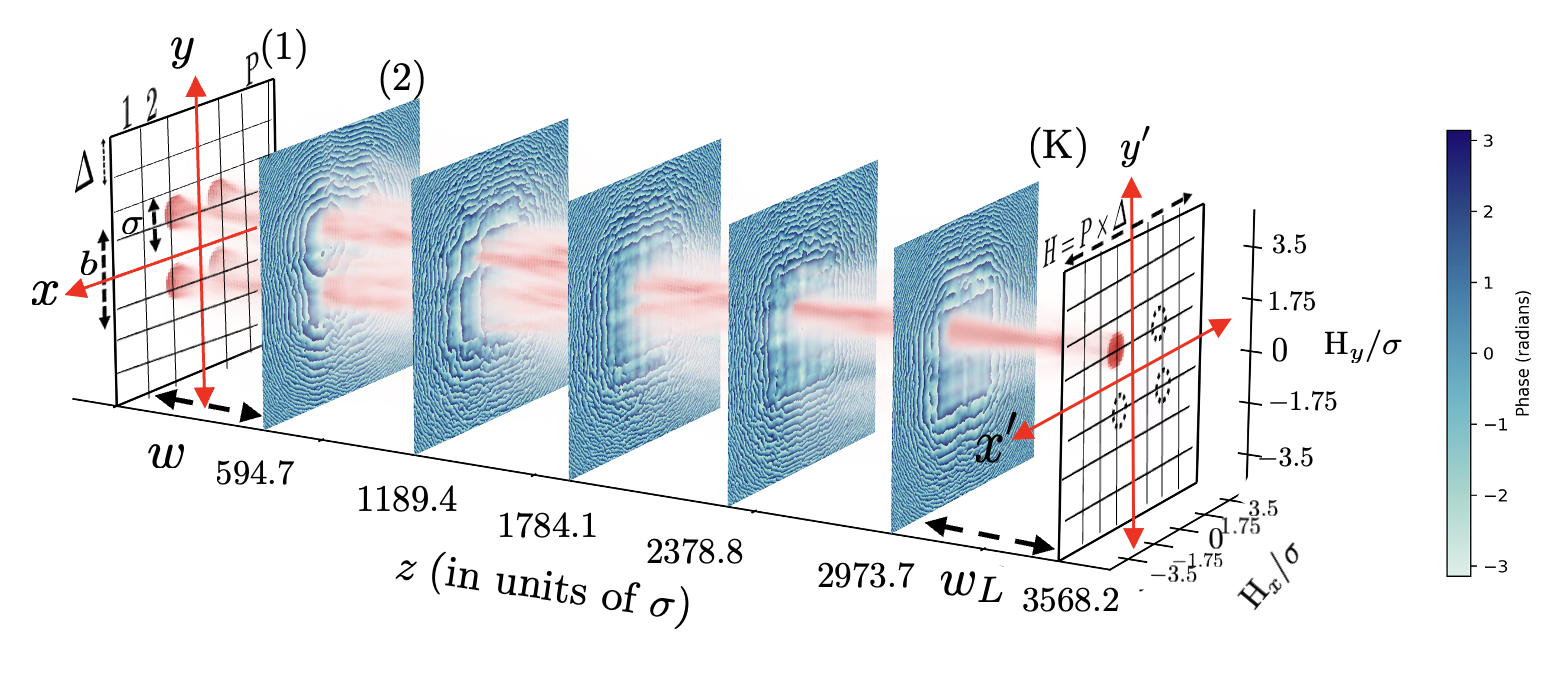}
    \caption{Simulated beam propagation through a $5$-layer MPLC when implementing a $4 \times 4$ Hadamard unitary, shown to transform the input $|\alpha\rangle|\alpha\rangle|\alpha\rangle|\alpha\rangle$ into an output $|2\alpha\rangle|0\rangle|0\rangle|0\rangle$. The field envelope's $x$ and $y$ dimensional extents ${H}_x = {H}_y \equiv H$, defined as the transverse square-region with $99\%$ total-power confinement, is roughly $6\sigma$ to $6.5\,\sigma$ throughout the $K$ planes.}
    \label{fig:bulge}
\end{figure}

The phase masks are optimized using the wavefront-matching algorithm~\cite{ARocha2025,fontaine2019}, after re-interpreting the pixel-mode unitary as a spatial-mode sorter of a set of orthonormal modes $\left\{\zeta_m(x,y)\right\}$, $1\le m \le M$, at the input, each of which are superpositions of the input pixel modes: 
\begin{equation}
\zeta_m(x,y) \equiv \sum_{n=1}^M u_{mn} ^\ast\phi_n(x,y),
\end{equation}
with $u_{ij}$ being the $(i,j)$-th entry of the complex-valued unitary matrix $U$, into the individual output pixel modes $\left\{\psi_m(x',y')\right\}$. See Appendix~\ref{app:unitary_as_modesorter} for the complete description of how this translation is performed. The iterative wavefront-matching (WFM) algorithm alternates between forward and backward propagations of superpositions of input pixel modes and the output pixel modes while updating the spatial phase profiles at each plane to minimize the error between the desired and realized unitary transformations. See Appendix~\ref{app:wavefrontmatching} and ~\ref{app:wavefrontmatchingconv} for further details of the WFM algorithm and convergence.

\section{Simulations and scaling analysis}~\label{sec:simulations}

We numerically evaluate our MPLC architecture to realize various classes of unitary transformations on 2, 4, 8, and 16 pixel modes, which includes the following:
\begin{itemize}
\item {\em Tunable-transmissivity two-input beamsplitter}---which has a variety of use cases ranging from a low-loss variable coupler interfering light from VCSEL arrays, to realizing high-visibility Hong-Ou-Mandel like interferences between photons emitted from single-photon emitters, 
\item {\em Hadamard transformations}---for use in receiver designs for super-additive communications capacity~\cite{Guha2011}, and efficient loading of weak-thermal starlight into quantum-memory registers~\cite{Khabiboulline2019})
\item {\em Haar random unitaries}---for possible use in realizing a highly-scalable Boson Sampling~\cite{Aaronson2011} or Gaussian Boson Sampling demonstration~\cite{Hamilton2017},
\item {\em Spatial permutations}---inspired by use in high-speed interconnects in high-performance computing clusters~\cite{Shang2022},
\item {\em Boosted fusion unitaries}---for use in realizing single-photon-ancillas assisted boosting of linear-optic Bell-State Measurement (BSM) success probability for dual-rail photonic qubits beyond $50\%$~\cite{Ewert2014}, and 
\item {\em Partial unitaries} acting on subsets of modes---inspired by use in heralded assembly of photonic cluster states via small resource cluster states and two-qubit fusion gates, assisted by percolation~\cite{Pant2019}.
\end{itemize}

We evaluate performance using the following two metrics (see Appendix~\ref{app:definitions} for the formal definitions):
\begin{enumerate}
\item {\em Loss}---We define the average transmissivity $\bar{\eta}$ to be the fraction of the energy in the $\zeta_m$-only excited input field that is collected across all the output pixel modes; averaged over all $m \in \left\{1,\ldots,M\right\}$.
\item {\em Crosstalk}---The average crosstalk $\bar{\epsilon}$ is defined as the fraction of the light intensity collected across all $M$ output pixel modes when the $\zeta_m$ input mode is excited that goes into the incorrect output pixel modes; averaged over all $m \in \left\{1,\ldots,M\right\}$.
\end{enumerate}

\subsection{Transmissivity and crosstalk}

\begin{figure}
    \centering
    \includegraphics[width=\columnwidth]{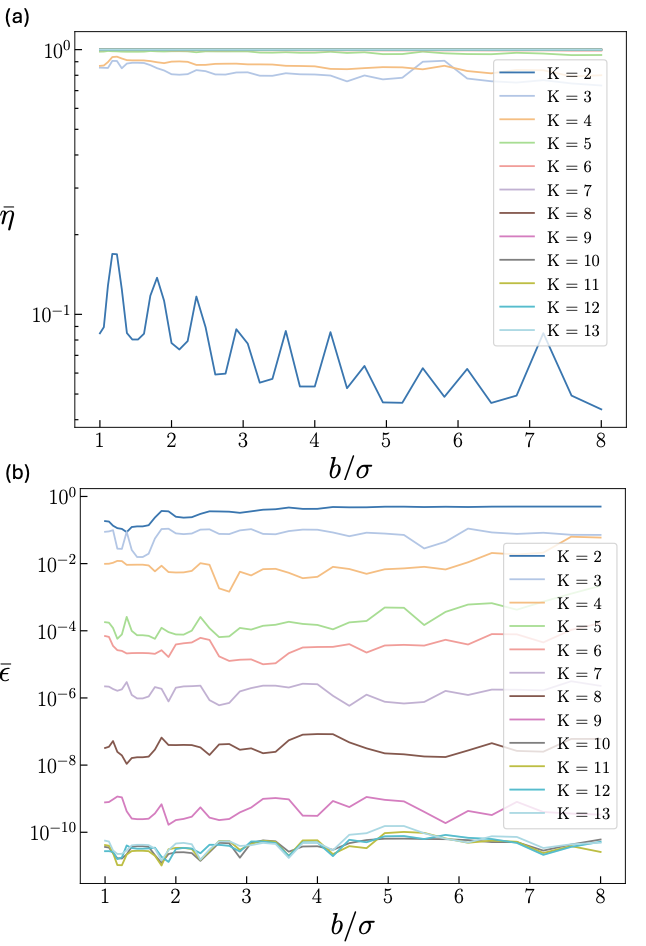}
    \caption{(a) Mean transmissivity $\bar{\eta}$ and (b) mean crosstalk ${\bar \epsilon}$, plotted as a function of MFD-normalized inter-pixel-mode spacing $b/\sigma$, for the 50-50 beamsplitter unitary in Eq.~\eqref{eq:beamsplitter}.}
    \label{fig:performance_beamsplitter}
\end{figure}
In this subsection, we report numerical results for the average transmissivity and average crosstalk, as a function of inter-pixel-mode spacing $b/\sigma$, for various unitaries discussed at the beginning of Section~\ref{sec:simulations}. 

As shown later in the subsection~\ref{sec:pixelation}, the transmissivity remains stable while crosstalk remains acceptably low for SLM pixel size $\Delta/\sigma \approx 0.3$ or lower, so we will use $\Delta = 0.3\sigma$ for our simulations. Further, based on the discussion in subsection~\ref{sec:planecountscaling}, we will use plane spacing of \( w/\sigma = 595\). This choice ensures optimal (lowest) slope of the $K \propto M$ scaling discussed in subsection~\ref{sec:planecountscaling}. Finally, we will assume center wavelength $\lambda= 532~\text{nm}$ for our simulations, and our mode-field diameter , $\sigma = 350 \mu m$, will be held fixed throughout our simulation unless otherwise specified. We also fix the SLM transverse length ${H}= 35\times \sigma$ for all the simulations; this value is sufficiently high to ensure no significant intensity is lost due to SLM-edge induced curtailment of the propagating field. 

Fig.~\ref{fig:performance_beamsplitter} plots $\bar{\eta}$ and $\bar{\epsilon}$ versus $b/\sigma$, for $K = 2, \ldots, 13$, for the $2$-mode 50-50 beamsplitter whose unitary is given by 
\begin{equation}
U = \frac{1}{\sqrt{2}}\left(\begin{array}{cc}
1 & 1 \\
1 & -1
\end{array}\right).
\label{eq:beamsplitter}
\end{equation}
We see that $\bar{\eta}$ and $\bar{\epsilon}$ both plateau for $b/\sigma \gtrsim 2$. 

In Appendix~\ref{app:examples}, we show similar performance evaluation results for: 
\begin{enumerate}
\item An alternative symmetric 2-mode beamsplitter that plays a role in several important circuits that appear in linear-optical quantum computing (see Fig.~\ref{fig:performance_altbeamsplitter});
\item Hadamard unitaries for growing size (see Fig.~\ref{fig:performance_16_mode_hadamard} for the results for the $16$-mode real Hadamard unitary arranged in a $4 \times 4$ pixel-mode grid);

\item Passive unitary circuits that appear in realizing single-photon ancilla-assisted boosted Bell State measurements on dual-rail photonic qubits (see Figs.~\ref{fig:performance_boostedBSM8}
 and~\ref{fig:performance_boostedBSM16}); and 

\item Unitaries on a partial set of pixel modes (i.e., while performing identity transformations on the remaining modes), inspired by fusion-based photonic cluster state preparation circuits~\cite{Pant2019}. See Fig.~\ref{fig:partialbeamsplitters}.
\end{enumerate}

In these examples, we see that $b/\sigma = 2$ is a reasonably good threshold value for designing our pixel-mode unitary architecture regardless of the value of $M$ (the size of the unitary), i.e., for $b/\sigma < 2$, the performance starts deteriorating. Therefore, henceforth, for all simulations, we will set $\sigma = 350 \mu$m and $b/\sigma = 2$. The qualitative behavior of the convergence of performance ($\bar \eta$ and $\bar \epsilon$) with increasing $K$ and $b/\sigma$ appear similar for each of these simulations. The above said convergence appears slower for the case of partial beamsplitters. Understanding the rigorous dependence of this convergence on the structure of the unitary is left open for future research.

\subsection{Impact of phase mask pixelation}~\label{sec:pixelation}

\begin{figure}
    \centering
    \includegraphics[width=\columnwidth]{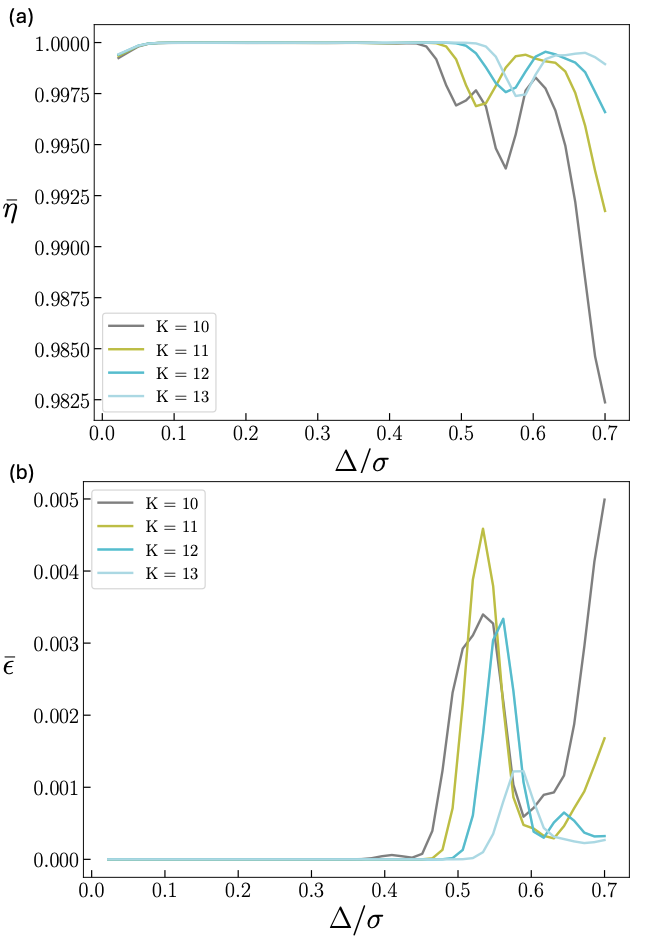}
    \caption{(a) Mean transmissivity $\bar{\eta}$ and (b) average crosstalk $\bar{\epsilon}$ versus pixel dimension $\Delta$, for the $2$-pixel-mode $50$-$50$ beamsplitter, with $K$ $= 10$ to $13$.}
    \label{fig:transmissivity_delta}
\end{figure}

For general universal convergence~\cite{morizur2010}, the pixel size $\Delta$ must become vanishingly small. To capture the effect of the pixel size to our architecture, and to inform pixelation choice in hardware, we evaluate and plot in Fig.~\ref{fig:transmissivity_delta}, the average transmissivity $\bar{\eta}$ and the average crosstalk $\bar{\epsilon}$ for increasing $\Delta/\sigma$ (pixel width normalized to the mode field diameter), respectively, for plane count $K=10,11,12$ and $13$. Both metrics are seen to remain essentially constant for $\Delta/\sigma \lesssim 0.3$, indicating robust performance in this regime. However, beyond $\Delta/\sigma \approx 0.3$, transmissivity begins to drop noticeably, and crosstalk rises sharply, with performance degradation becoming more pronounced near $\Delta/\sigma \approx 0.5$--$0.7$. This trend is consistent across all tested values of $K$ and across the larger unitaries considered here; although higher plane counts mitigate the severity of the drop. Based on these observations, we will fix the pixel size to $\Delta = 0.3\,\sigma$ for all subsequent simulations.

\subsection{Plane count scaling}\label{sec:planecountscaling}

In the subsection we will report on our (numerical) finding that achieving fixed performance targets requires that the number of phase planes $K$ scales linearly with the number of modes $M$. Closely intertwined with this analysis is the optimization of the inter-plane Rayleigh-Sommerfeld propagation length $w$. Since we need a family of unitaries with growing $M$, we choose the complex Hadamard unitary, with matrix elements:
\[
T_{mn} = \frac{1}{\sqrt{M}}e^{i 2\pi mn/M}, |T_{mn}| = \frac{1}{\sqrt{M}}, \quad TT^{\dagger} = {\mathbb I}_{{M}},
\]
i.e., all entries have equal magnitude ${1}/{\sqrt{{M}}}$ and the rows/columns are orthonormal vectors. 

\begin{figure*}
    \centering
    \includegraphics[width=0.9\textwidth]{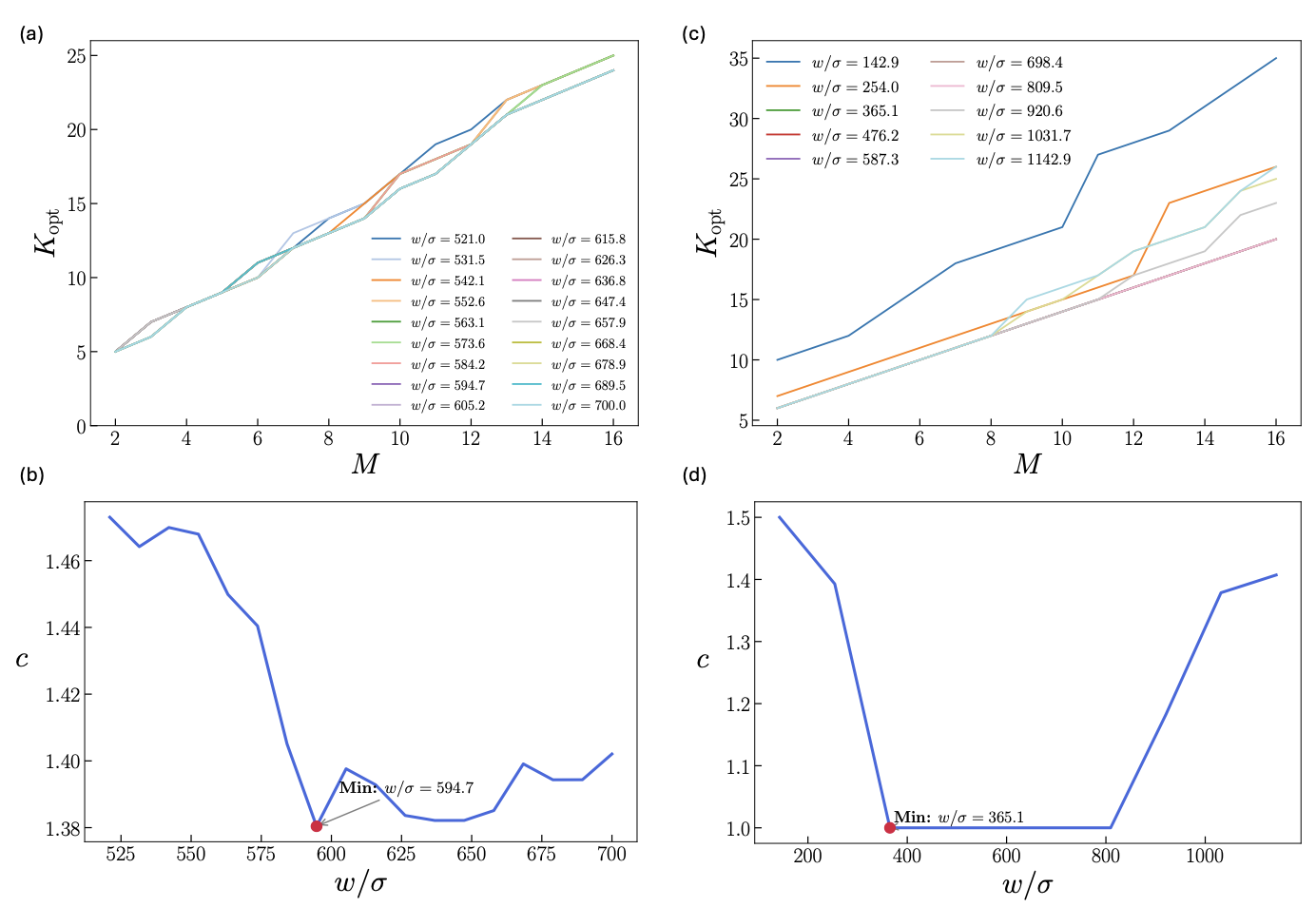}
    \caption{
    {\em Left panel:} (a) Required plane count $K$ versus mode count $M$ to realize an $M$-mode complex-Hadamard unitary at fixed $\bar{\eta}=0.95$ and $\bar{\epsilon}=0.005$, shown for different plane spacings $w/\sigma$. 
    (b) Linear fit $K=cM$ to the numerically computed $K$, with proportionality constant $c$ plotted as a function of $w/\sigma$. 
    {\em Right panel:} (c) Same as (a) but for an $M$-mode random-permutation unitary. 
    (d) Corresponding proportionality constant $c$ versus $w/\sigma$.}
    \label{fig:plane_scaling}
\end{figure*}
Figure~\ref{fig:plane_scaling}(a) plots the plane count $K$ needed to achieve ${\bar \eta} = 0.95$ and ${\bar \epsilon} = 0.005$, to realize the $M$-mode complex-Hadamard pixel-mode unitary, evaluated for various plane-spacings, $w/\sigma$. The lower envelope of these plots, termed $K_{\rm opt}(M)$---the minimum plane count---is seen to scale roughly linearly with $M$. For each value of $w/\sigma$, we fit this numerically-computed plane count $K$, to $K = cM$, extract the proportionality constants $c$, and plot it as a function of $w/\sigma$, in Figure~\ref{fig:plane_scaling}(b). The plot shows a minimum at roughly, $w/\sigma \approx 595$, suggesting a favorable trade-off between diffraction spreading and inter-mode separation: at this value, fewer planes are required to achieve a given $M$. For all of the above simulations, we held ${H}/\sigma = 35$---picked to ensure that as we increases $w$ there was no significant field-clipping by the SLM plane boundaries due to diffraction spread---and held the pixel size $\Delta/\sigma = 0.3$, pertaining to our findings in Section~\ref{sec:pixelation}.

We repeat the above numerical calculation for a different growing class of unitaries: where $U$ is a random permutation matrix, $D_{mn} = \delta_{n, \pi(m)}$, and report the corresponding numerical findings in Figs.~\ref{fig:plane_scaling}(c) and (d). The qualitative findings are the same: $c$ drops sharply as the inter-plane spacing $w/\sigma$ is increased, and saturates to a rough plateau starting at a particular value of $w/\sigma$. This roughly-optimal $w/\sigma \approx 595$ for complex Hadamard unitaries, and $\approx 365$ for random-permutation unitaries. 

Once the plane spacing has been optimally chosen,
\begin{equation}
K_{\rm opt} \approx c M,
\end{equation}
where $c$ only would depend on the target $({\bar \eta},{\bar \epsilon})$ and the class of unitary we wish to realize. Note that the circuit depth for a $M$-mode unitary in Clements {\em et al.}'s MZI mesh is $\propto M$ as well. In the original MPLC paper by Morizur {\em et al.}, the universality of the MPLC based $M$-mode unitary was proven by realization of $2$-mode MZI-like unitaries on pairs of modes (while performing identity on the remaining $M-2$ modes), which required a constant number of planes; which was then followed by realizing $M(M-1)/2$ MZIs (in a sequence) whilst breaking up $U$ using the Clements {\em et al.} algorithm. This resulted in $O(M^2)$ MPLC planes to realize a general $M$-mode unitary. Our calculations show that the MPLC is more powerful than that, and if used in its native form, needs $O(M)$ planes. Solutions to SLM phases to realize a unitary using a $K$-plane MPLC is highly degenerate. 

It is likely that the constant $c$ in the above scaling can be improved (lowered) by a better MPLC compilation than the WFM algorithm we used, e.g., by optimizing the geometry and shapes of the sorted regions in the output plane, or using machine learning rather than the forward-backward propagation method to find the SLM phases~\cite{Bearne2026}.

\subsection{Stackability and field bulge}

\begin{figure}
    \centering
    \includegraphics[width=\columnwidth]{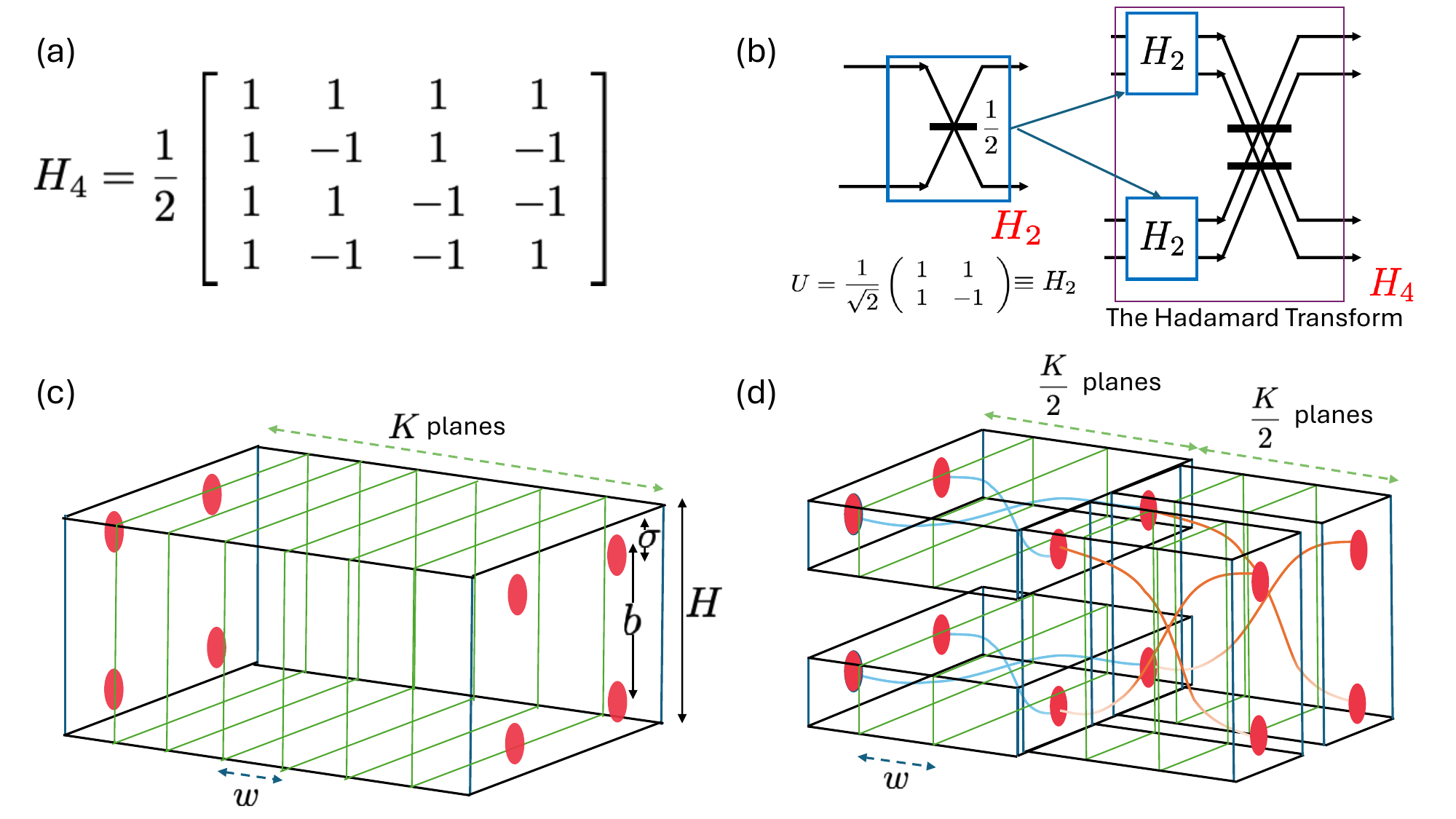}
    \caption{(a) The $4$-mode Hadamard unitary $H_4$; (b) The Green Machine architecture for realizing the Hadamard unitary using 4 50-50 beamsplitters; (c) A {\em direct} MPLC realization of $H_4$ via $K$ phase masks; (d) A {\em stacked} realization of $H_4$ where two 50-50 beamsplitters are realized in parallel using a $K/2$-plane MPLC, followed by a parallel realization of another pair of 50-50 beamsplitters using another $K/2$-plane MPLC.}
    \label{fig:stacking}
\end{figure}
In Fig.~\ref{fig:stacking}(a), we write the expression of the $4$-mode real-Hadamard unitary, and in Fig.~\ref{fig:stacking}(b), we show its representation as two columns of two beamsplitters each. This particular {\em Green Machine} realization of the $4$-mode Hadamard unitary, originally proposed by Guha~\cite{Guha2011}, proved particularly useful in its time-bin realization, i.e., when across four time-separated temporal `pixel modes' (or pulse slots), by Cui {\em et al.}~\cite{Cui2025} because of the fact that the two beamsplitters in each column could be realized by one physical beam splitter each, by a careful sequence of time delays applied to the entire pulse sequence. In our spatial-pixel-mode realization, it turns out that the Green Machine realization performs significantly worse compared to the direct WFM-compiled realization of $H_4$. 

\begin{figure*}
    \centering
    \includegraphics[width=0.9\textwidth]{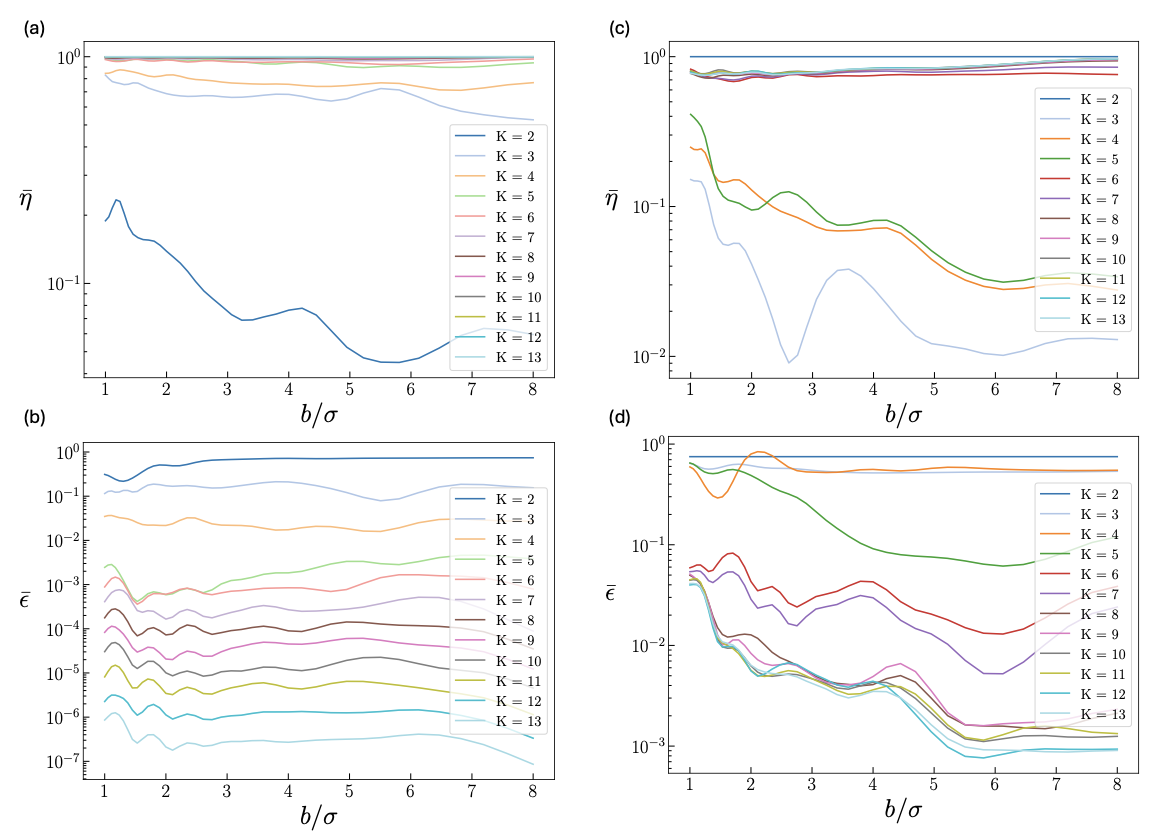}
    \caption{{\em Left panel}: (a) Transmissivity and (b) crosstalk for the stacked-realization of $H_4$, as a function of $b/\sigma$ (mode separation in unit of mode-field diameter), for $K \in \left\{2, \ldots, 13\right\}$; {\em Right panel}: (c) Transmissivity and (d) crosstalk for the direct-realization of $H_4$, as a function of $b/\sigma$.}
    \label{fig:stacking_results}
\end{figure*}
In Fig.~\ref{fig:stacking_results}, we plot the average transmission efficiency $\bar{\eta}$ and the average crosstalk $\bar{\epsilon}$ as functions of $b / \sigma$ (spacing between adjacent pixel modes measured in units of the mode-field diameter of one pixel mode). We also include results for directly encoding a $4 \times 4$ unitary in the MPLC. These simulations assume ${H}/\sigma= 35$, $\Delta/\sigma= 0.3$ and $w/\sigma= 595$. Direct encoding of $H_4$ significantly outperforms the stacking-based beamsplitter approach in both transmissivity ($\bar{\eta}$) and crosstalk ($\bar{\epsilon}$), by a few orders of magnitude for crosstalk (see the $K=4$ plots for instance). 

There are two reasons for this improved performance: first is the direct compilation does not constrain the MPLC to compile the circuit into specific two-mode beamsplitters (which is not optimal for its native functionality). Second, when realizing  a pixel-mode unitary using an MPLC, the field---while propagating through the planes---first diverges and then collects itself back into the output plane pixel modes (see Fig.~\ref{fig:bulge}). This {\em field bulge} can be much larger compared to the inter-pixel-mode spacing in the transverse plane, and hence if two unitaries are realized in parallel using the same SLM planes (e.g., the pairs of 2-input beamsplitters realized in parallel in the stacked configuration shown in Fig.~\ref{fig:stacking}(d)), the native field bulges of each unitary can spoil the other's functionality.

\begin{figure}
    \centering
    \includegraphics[width=\columnwidth]{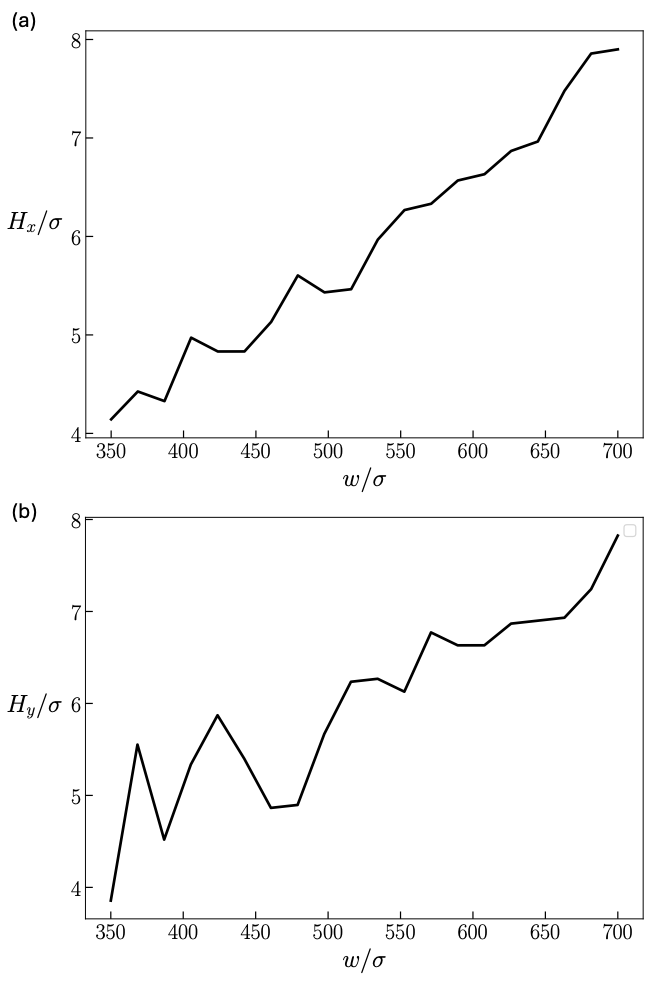}
    \caption{The {\em field bulge}, parametrized by $H_x$ and $H_y$---maximum transverse spread of the field-in-propagation such that $99\%$ of the field's transverse intensity profile is contained, throughout the $K$-plane propagation---plotted as function of plane spacing $w/\sigma$, for the $M=4$ complex Hadamard unitary. We hold $b=2\sigma$, and $\Delta = 0.3\sigma$.}
    \label{fig:bulge_vs_planespacing}
\end{figure}
To quantify the above further, we consider the $M=4$ complex Hadamard unitary and plot the field bulge in both the $x$ and $y$ directions as a function of $w/\sigma$, in Fig.~\ref{fig:bulge_vs_planespacing}. We define ${H}_x$ and ${H}_y$ as the lengths of the field bulge along the $x$- and $y$-directions such that $99\%$ of the field's transverse intensity profile is contained within this length throughout the $K$-plane propagation. The field bulge is seen to increase with $w / \sigma$, which is expected due to diffraction spread. But, most importantly, we see that at the MPLC-performance-optimum value of $w/\sigma \approx 595$, both $H_x$ and $H_y$ are roughly $6\sigma$, which is much larger than the inter-pixel-mode transverse spacing of $b = 2\sigma$ that we have taken for all our calculations. This explains the field-bulge-induced crosstalk that deteriorates the performance of the stacked realization in Fig.~\ref{fig:stacking}(d).

\section{Experimental demonstration}~\label{sec:experiments}

In this section, we will describe our programmable pixel-mode unitary realization in practice. The input spatial field is generated using a coherent spatial scene generator implemented with an SLM and a 4$f$ imaging system. The MPLC itself is realized by reflecting the optical field between the SLM and a mirror, with different regions of the SLM encoding successive phase planes. The output field is measured using an EMCCD camera. All experiments were performed using a continuous-wave laser at a wavelength of 637 nm. We experimentally demonstrate:
\begin{itemize}
\item Programmable beamsplitters between two pixel modes
\item Hadamard transformations on 4, 8, and 16 modes
\item Spatial permutations between modes
\item Partial unitaries acting on subsets of modes
\end{itemize}

\subsection{Experimental Setup}
  \begin{figure*}[t]
     \centering
    \includegraphics[width=1\linewidth]{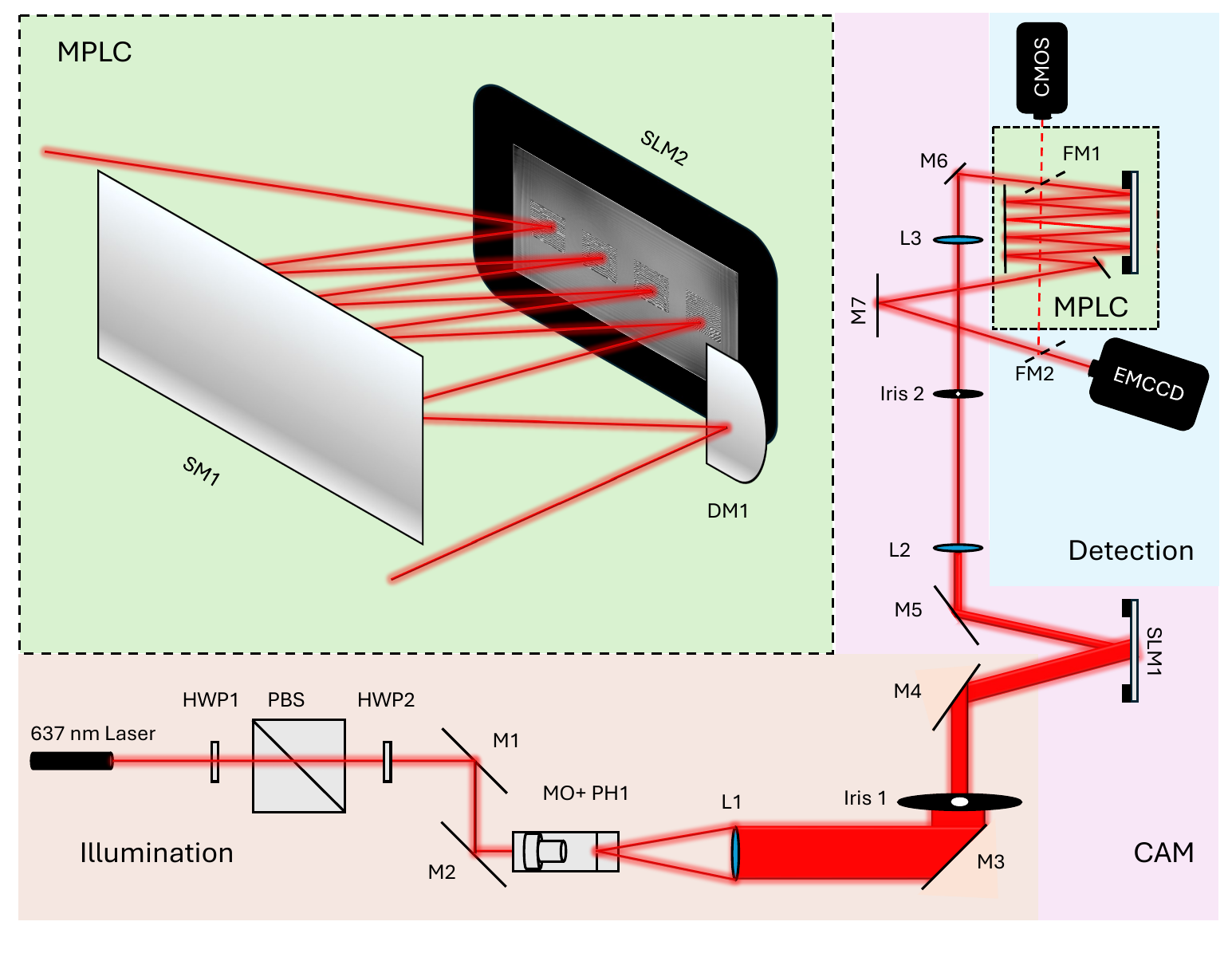}
    \caption{ Experimental realization of complex-amplitude encoding and multi-plane light conversion (MPLC).
A 637-nm laser is power- and polarization-controlled (HWP1, PBS, HWP2), spatially filtered (Microscope objective-MO, PH1), and expanded to illuminate SLM1, where the target complex field is encoded. A $4f$ relay with an iris images the encoded object onto the MPLC stage. The MPLC is implemented on SLM2 in a folded multi-pass geometry (with square mirror SM1), enabling four sequential phase modulations. The transformed field is recorded on an EMCCD for crosstalk measurements, while flip mirrors (FM1, FM2) redirect the beam to a CMOS camera for input–output power calibration and transmissivity measurements.}
    \label{fig:System_setup}
\end{figure*}
The schematics of our experimental apparatus is shown in Fig.~\ref{fig:System_setup}. The system consists of four functional subsystems: (i) illumination and state preparation, (ii) complex-amplitude modulation (CAM) encoder, (iii) multi-plane light conversion (MPLC), and (iv) detection and calibration.
\begin{enumerate}[label=(\roman*)]
    \item \textbf{Illumination:} A continuous-wave laser diode at a center wavelength of 637 nm (Edmund Optics, 50 mW USB laser) serves as the light source. The output beam first passes through a half-wave plate (HWP1) and a polarizing beam splitter (PBS), enabling continuous control of the optical power coupled into the system. A second half-wave plate (HWP2) sets the polarization state required for phase-only modulation at the spatial light modulators (SLMs).

    The beam is subsequently spatially filtered and collimated using a microscope objective (MO, Newport M-10X), a 5-µm pinhole (PH1, Newport 910PH-5), and a 100-mm focal-length lens (L1) paring with an iris (Iris 1). This stage produces a illumination beam in a clean Gaussian beam profile and sends it to the following complex amplitude modulation stage for wavefront encoding. 
    
    \item \textbf{Complex amplitude modulation (CAM):} The collimated beam is directed via a mirror (M4) to the first spatial light modulator (SLM1, Holoeye Pluto-2.1-NIR-145) at an angle less than five degrees. We use the method described in Ref.~\cite{arrizon2007CAM} to transfer the phase-only modulation mask displayed on the SLM1 onto the incoming wavefront of the illumination beam, ensuring the generation of the desired complex field patterns at the output of the 4f system.

    After reflection from SLM1, the beam propagates through a $4f$ imaging system (formed by L2 and L3) with a spatial filter (Iris 2) placed at the Fourier plane to suppress unwanted diffraction orders and residual unmodulated light and to ensure that only the encoded object field is transmitted into the following MPLC stage. The $4f$ system also relay the generated complex field  directly onto the first MPLC plane, providing precise mode matching and spatial registration. This is an essential requirement for demonstrating high-fidelity MPLC.
    
    \item \textbf{MPLC:} The MPLC stage is implemented using a second SLM (SLM2, Holoeye Pluto-2.1-NIR-145) in a folded multi-pass configuration (see Fig.~\ref{fig:System_setup}). After the light reaching SLM2, the beam undergoes multiple reflections between the SLM and a unframed square mirror. Each time the light reflects at a different section of the SLM, which applies pixel-wise programmable phases -- a phase mask -- to the wavefront of the light, equivalent to an inline MPLC with a gap of 240 mm between phase planes. Thus, this MPLC setup effectively realizes the optimized design shown in Fig.\ref{fig:bulge}.

    Each MPLC phase plane section occupies a 300 $\times$ 300 pixel region (2.4 mm $\times$ 2.4 mm) on the SLM, with equal spacing between adjacent planes to prevent overlap. In addition to the designed phase profile, a blazed grating (8-pixel period) is superimposed on each plane. The grating deflects the modulated beam away from specular reflection, ensuring that only light that has interacted with the programmed phase mask remains in the optical path. Unmodulated or straying light fields are therefore isolated from the rest of the system, greatly reducing the noise seen by the camera.

    Unless otherwise specified, all demonstrations and trend studies were performed under the following system parameters. The illumination wavelength was $\lambda = 637\mathrm{nm}$, and the SLM pixel pitch was $\Delta = 8\mu\mathrm{m}$. Pixel modes were encoded as Gaussian beams (HG$_{00}$) with mode-field diameter $\sigma = 200\mu\mathrm{m}$, defined at the $1/e^2$ intensity radius. Adjacent pixel modes were separated by $b = 3\sigma$ to ensure negligible overlap at the input plane. 

    The transmissivity of each phase plane in MPLC is measured to be 2.0 dB to 2.5 dB for different phase masks implemented at the SLM2. This loss includes the 1 dB loss of the raw reflectivity of the SLM pixels, the SLM filling factor of $93\%$, and the measured blaze grating efficiency of $76-83\%$ when overlapping with the designed phase masks.

    %The optical path length between successive planes and between the final plane and the detection stage is measured and incorporated into the phase-mask design to ensure correct free-space propagation between planes.
    
    \item \textbf{Detection:} The transformed wavefront is then directed to an EMCCD camera (Andor iXon Ultra 897) placed at a distance of $w_L = 840\mathrm{mm}$ away from the final MPLC plane. The EMCCD counts the photon arrived at each pixel with the detection efficiency of $\sim90\%$ and records the output intensity distribution at the wavefront.

    We also implemented a pair of flip mirrors (FM1 and FM2) for quick and repeatable characterization of the transmissivity of each designed MPLC, to redirect the beam to a CMOS camera (Thorlabs CS235MU). FM1 enables measurement of the input intensity distribution prior to the MPLC stage, while FM2 enables measurement of the total output intensity after the MPLC. Comparing these calibrated measurements while backing out aforementioned SLM losses allows the extraction of system transmissivity independent of modal crosstalk. 
\end{enumerate}

\begin{figure}
    \centering
    \includegraphics[width=\columnwidth]{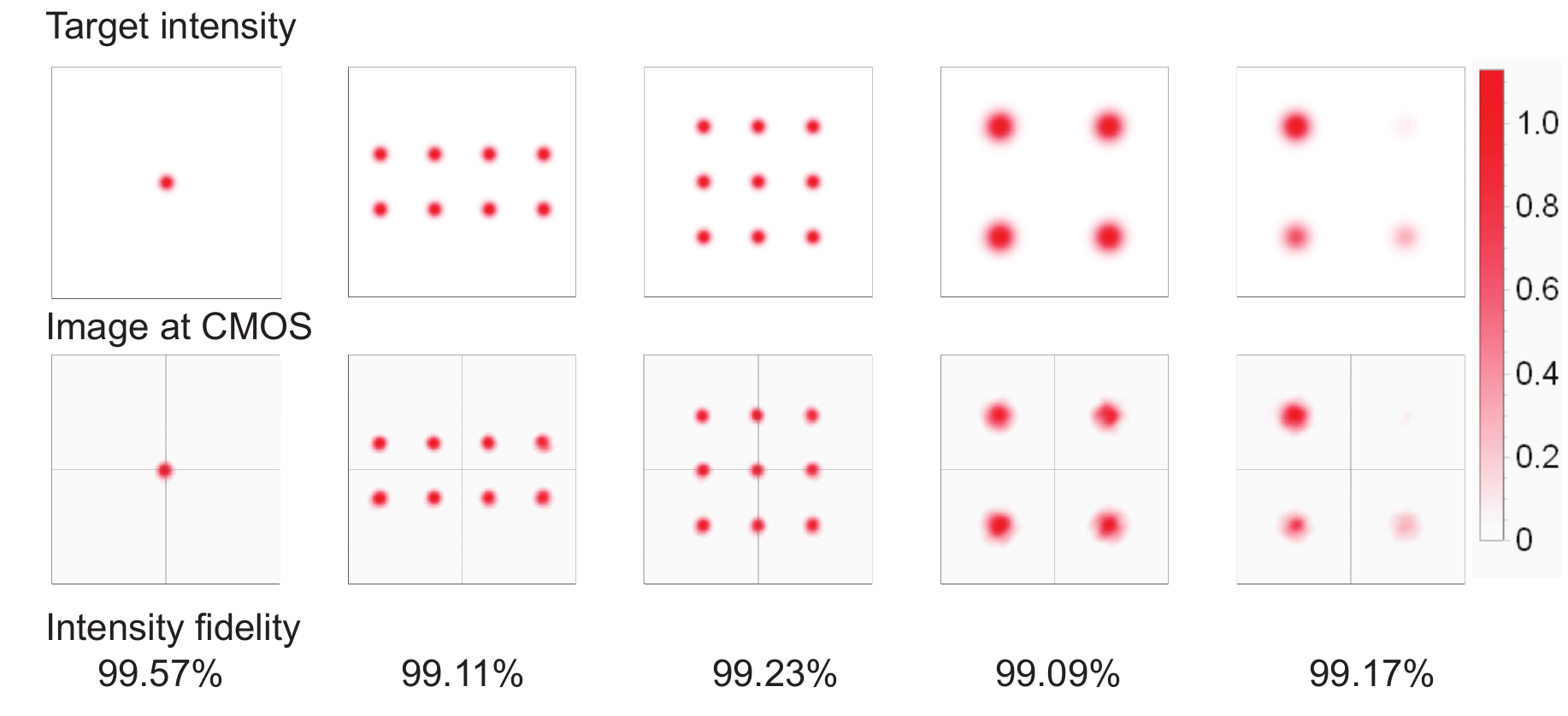}
    \caption{Verification of high-fidelity generation of the desired coherent fields that are then imaged onto the MPLC based pixel-mode unitary.}
    \label{fig:CAMcharacterization}
\end{figure}
\begin{figure*}
    \centering
    \includegraphics[width=1\linewidth]{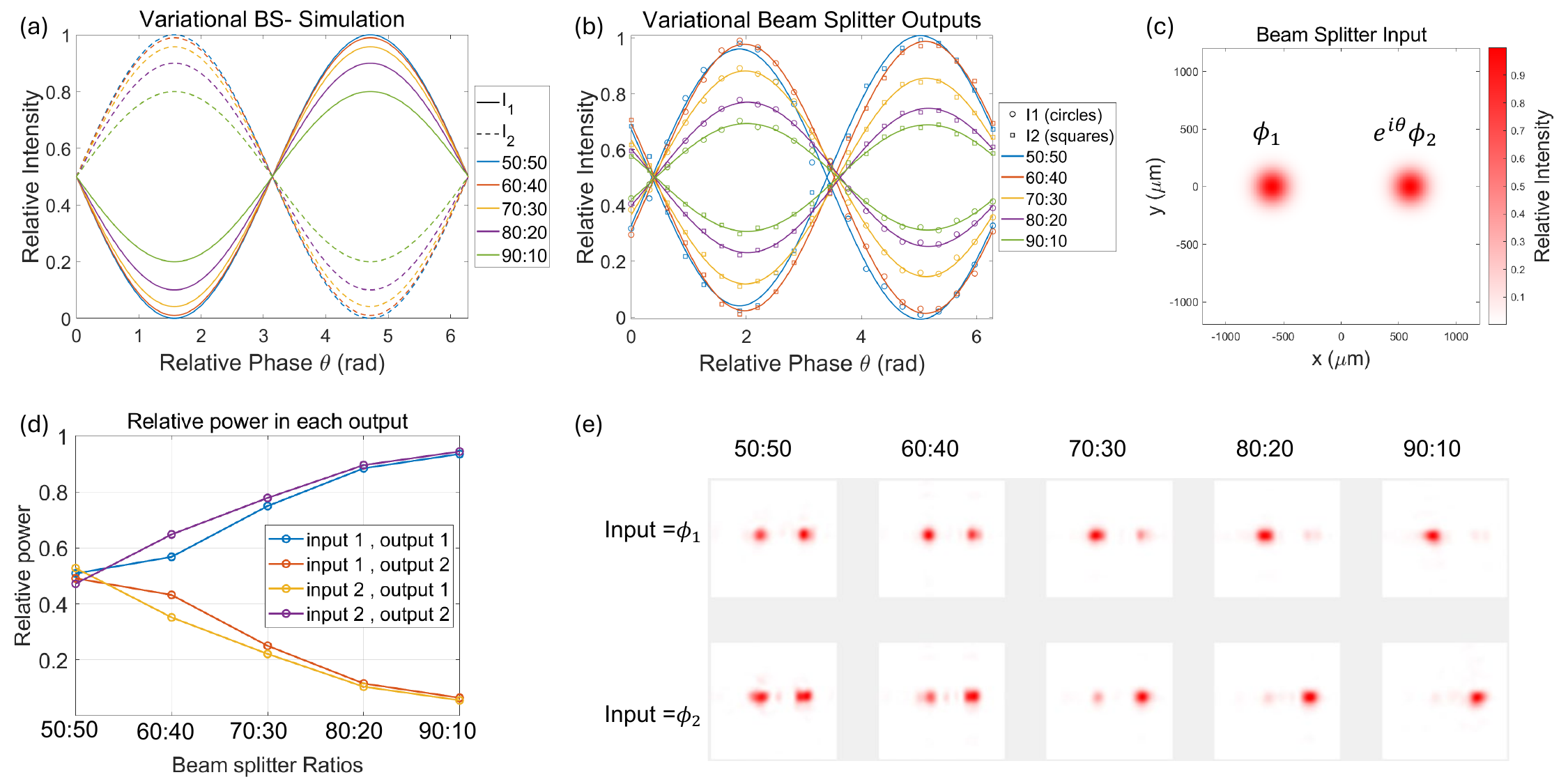}
    \caption{This figure shows measured performance of the MPLC programmed as a tunable two-mode beamsplitter: (a) shows the simulated relative intensity, $I_1$ and $I_2$ in each output mode as a function of the relative phase between the two input modes incident on beam splitters with different splitting ratios. (b) shows the relative intensity in each output mode as a function of the relative phase between the two input modes encoded by our complex amplitude encoding set up and incident on our MPLC as it is programmed to act as a beam splitter with different splitting ratio. (c) shows the simulated intensity distribution of the two input pixel modes in $1\times2$ tiling, separated by $b=3\sigma$. (d) shows the power at each designated output spot as a function of the MPLC's beam splitting ratios. For this demonstration, we used input state 1 ($\ket{\alpha}_{\phi_1}\ket{0}_{\phi_2}$) and input state 2 ($\ket{0}_{\phi_1}\ket{\alpha}_{\phi_2}$).  (e) shows the output EMCCD images for input state 1 and 2 respectively. }
    \label{fig:Tunable_Beam_Splitter}
\end{figure*}
\begin{figure*}
    \centering
    \includegraphics[width=\linewidth]{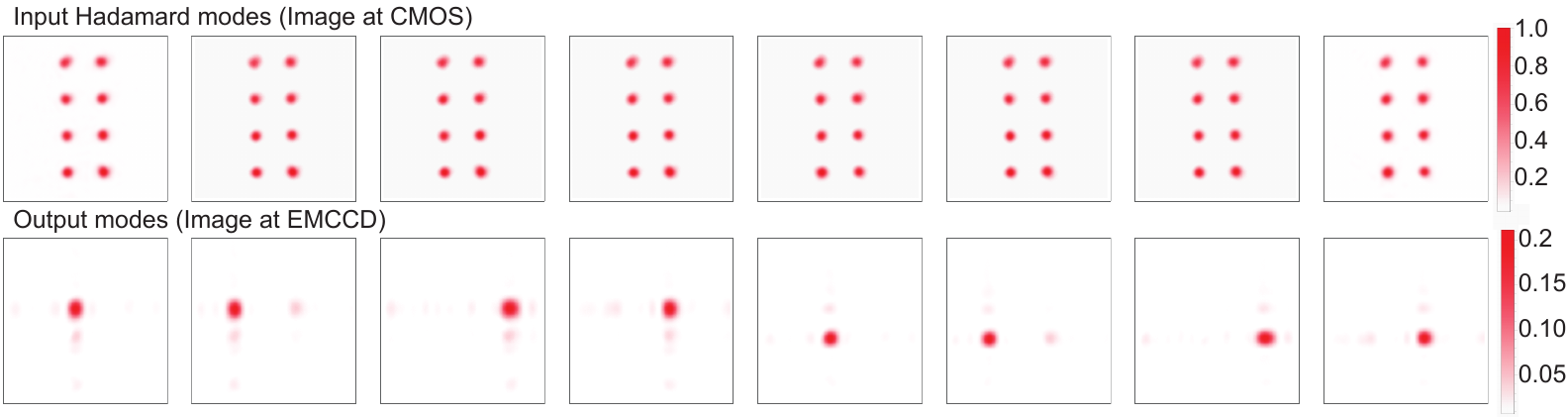}
    \caption{Realization of the $8$-mode Hadamard matrix, with the $8$ modes tiled in a $4 \times 2$ pattern ($b=3\sigma$). The top row shows the transmitted modes $\zeta_m(x,y)$, $1 \le m \le 8$, while the bottom row shows the intensity images of the received modes, which are non-ideal, yet high-fidelity, images of the corresponding output-plane pixel modes $\psi_m(x^\prime,y^\prime)$.}
\label{fig:Hadamard8experiment}
\end{figure*}
\begin{figure}
    \centering
\includegraphics[width=\linewidth]{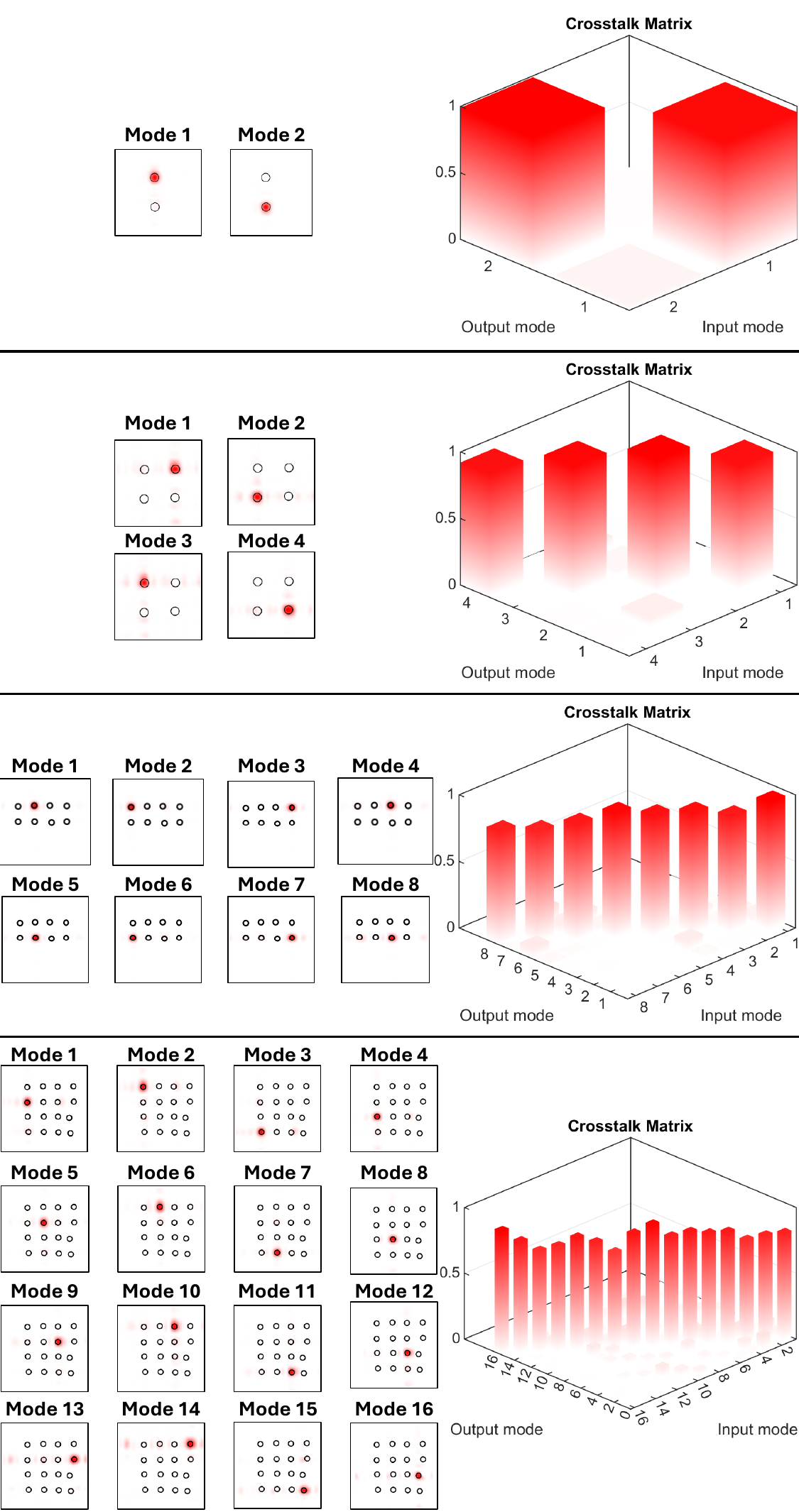}
    \caption{The measured intensity images of the output fields and the corresponding intensity crosstalk matrices for MPLC-realized Hadamard unitaries of dimensions $4$, $8$, and $16$. The input and output pixel modes' MFD for the $4$-modes Hadamard is $400\mu m$, and the input and output pixel modes' MFD for the $8$- and $16$- modes Hadamard is $200\mu m$.}
    \label{fig:hadamard4816}
\end{figure}

\begin{figure*}
    \centering
    \includegraphics[width=\linewidth]{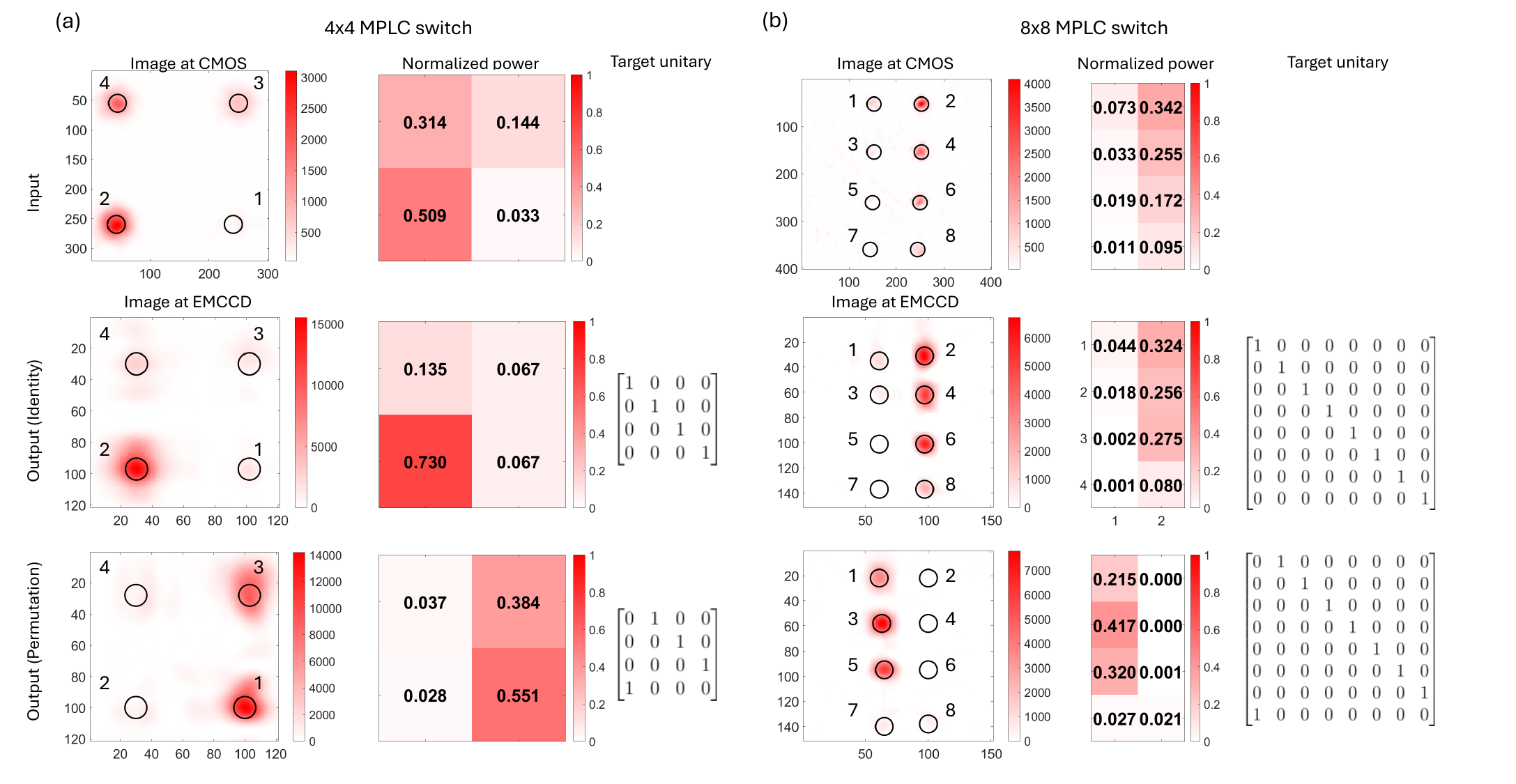}
    \caption{ MPLC as a reconfigurable optical switch acting on 4 (a) pixel modes and on (b) 8 pixel modes. The upper left images in (a) and (b) are the intensity distributions of the input field. After propagating through the MPLC system configured to perform the target unitaries, the output fields intensity distribution are captured by the EMCCD camera and shown respectively. The normalized power matrix quantifies the relative power in each pixel mode, with all values summing to 1. Tick marks indicate pixel numbers. The pixel pitches of the EMCCD and the CMOS are 16 $\mu$m, and 5.86 $\mu$m respectively.}
    \label{fig:optical_switch}
\end{figure*}
\begin{figure*}
    \centering
    \includegraphics[width=\linewidth]{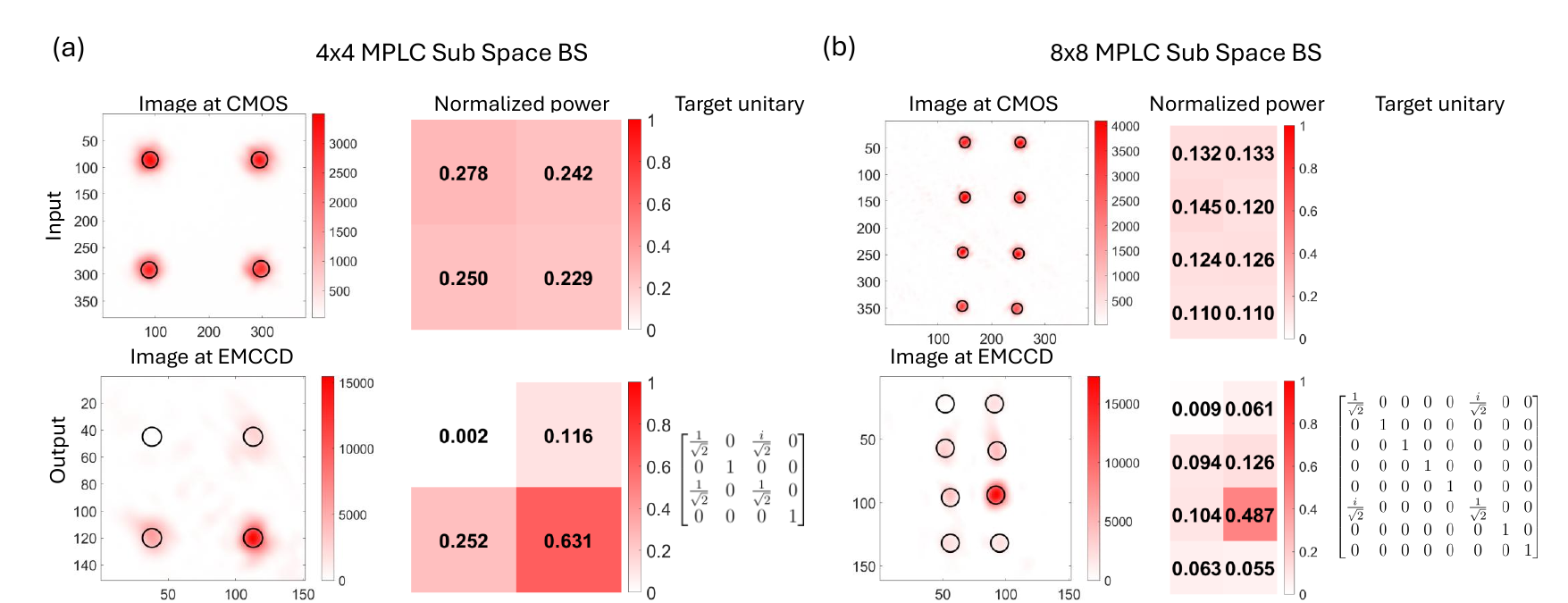}
    \caption{MPLC as a reconfigurable subspace beam splitter acting on 4 pixel modes (a) and 8 pixel modes (b). The upper left images in (a) and (b) are the intensity distributions of the input field. After propagating through the MPLC system configured to perform the target unitaries, the output fields intensity distribution are captured by the EMCCD camera and shown respectively. The normalized power matrix quantifies the relative power in each pixel mode, with all modes summing to 1. Tick marks indicate pixel numbers. The pixel pitches of the EMCCD and the CMOS are 16 $\mu$m, and 5.86 $\mu$m respectively.}
    \label{fig:subspace_BS}
\end{figure*}
\subsection{Results}
We first characterizes the CAM to ensure the performance of state preparation and encoding. The results measured at the CMOS camera and the comparison to the target intensity distributions are exhibited in Fig.~\ref{fig:CAMcharacterization}. All the intensity fidelities~\cite{rosales2017intensityfidelity} of tested patterns reach above $99\%$. The \textit{intensity fidelity} is the cross correlation of the image data to the corresponding discretized target mode intensity. 

After calibrating the CAM, we implement the designed phase masks in MPLC for realizing a two-mode beamsplitter with variable transmissivity. The input field is defined over two spatially separated HG$_{00}$ mode, $\phi_1$ and $\phi_2$, with a gap of $b=3\sigma$ in between, and the MFD is $\sigma=400\mu m$. We also use the CAM to encode different relative phase and intensity at the two mode, which effectively mimics any input superposition mode defined over this 2-dimensional Hilbert space spanned by the two modes. To characterize the beamsplitter with splitting ratio of $\eta:1-\eta$, we encode two equal-power coherent states into both modes as $|\alpha\rangle_{\phi_1}|\alpha e^{i\theta}\rangle_{\phi_2}$ with $\theta$ the relative phase to be swept and the notation $|\cdot\rangle_{\phi_{1,2}}$ labeling the occupied spatial mode. We let this input state passing through the MPLC, and check the intensity distributed at the two corresponding output modes on the EMCCD. The results are shown in Fig.~\ref{fig:Tunable_Beam_Splitter}(a) and (b). From the fitted data shown in Fig.~\ref{fig:Tunable_Beam_Splitter} (b), we obtained an average transmissivity of $58.97\%$ and visibility of $0.984$ for the $50:50$ beam splitter, which indicates the MPLC beamsplitter has potential for Hong-Ou-Mandel (HOM) interference between two quantum emitters emitting single photons into free space. We also test the performances of MPLC beamsplitter with splitting ratio $\eta$ of the tunable beamsplitter by shining one coherent state at only one of the input modes, i.e. $|\alpha\rangle_{\phi_1}$ or $|\alpha\rangle_{\phi_2}$. The relative power and intensity distributions of the output fields are shown in Fig.~\ref{fig:Tunable_Beam_Splitter} (d) and (e), respectively. The fidelity averaged over beamsplitters at the selected splitting ratios are 0.98.

For Hadamard unitary demonstration, we prepare coherent states in the effective sorting modes as the input of the MPLC system. We use pixel modes with $\mathrm{HG}_{00}$ spatial profile and separated horizontally and vertically with each other by $b=3\sigma$. 
Gaussian pixel modes are configured to have an MFD of 400 $\mu m$ for the 4-mode ($2\times2$ tile pattern) Hadamard unitary, and 200 $\mu m$ for the 8-mode ($4\times2$ tile pattern) and 16-mode ($4\times4$ tile pattern) Hadamard unitaries. We first image the input intensity pattern by the CMOS camera, and then image the output field after MPLC by the EMCCD camera. Fig.~\ref{fig:Hadamard8experiment} shows the measured intensity-only images of the input fields $\zeta_m(x,y)$ (top row) and the corresponding output fields of the target output-plane pixel modes $\psi_m(x',y')$ (bottom row), $1 \le m \le 8$ for the realization of the $8$-mode Hadamard unitary $H_8$. In Fig.~\ref{fig:hadamard4816}, we show the measured intensity images of the output fields and the corresponding crosstalk matrices obtained when implementing Hadamard unitaries of dimensions 4, 8, and 16. The measured average transmissivity are $\{69\%,87\%,69\%\}$ and the measured average crosstalk values are $\{5\%,10.9\%,17\%\}$ for the Hadamard unitaries of dimensions 4, 8, and 16 respectively. We note that the distance between the last phase mask and the EMCCD camera ($w_L$) was not chosen to optimize the loss and crosstalk performance for the entire range of MFDs we report results on. All our MPLC phase-mask designs were done targeting the output pixel-mode geometry and the MFD of the individual pixel modes to be identical to those at the input. However, $w_L$ was {\em not} optimized for each MFD individually. Our demonstrations therefore with the smaller MFDs such as $\sigma=200\mu m$, produced output pixel-mode sizes significantly larger than the input MFD of $200\mu$m. We expect the overall performance to improve when $w_L$ is chosen optimally, individually optimized for each MFD choice.

In Fig.~\ref{fig:optical_switch}, we test and show the capability of using MPLC to realize identity unitary transformation and a permutation unitary transformation for 4 pixel modes in panel (a) and 8 pixel modes in panel (b). For the four mode transformation, we prepare an input state $\ket{\alpha}_{\phi_1}\ket{4\alpha}_{\phi_2}\ket{2\alpha}_{\phi_3}\ket{3\alpha}_{\phi_4}$, with the intensity distribution shown in Fig.~\ref{fig:optical_switch} (a). After propagating through the MPLC system configured to perform the four-mode identity transformation and a four-mode permutation transformation respectively, the output fields' intensity distribution are recorded with the EMCCD camera and shown in Fig.~\ref{fig:optical_switch} (a). We quantify the relative power within each pixel modes with a normalized power matrix which has the total relative power in all modes sum up to one. Similarly for the eight-mode transformation, we prepare an input state $\ket{4\alpha}_{\phi_1}\ket{8\alpha}_{\phi_2}\ket{3\alpha}_{\phi_3}\ket{7\alpha}_{\phi_4}\ket{2\alpha}_{\phi_5}\ket{6\alpha}_{\phi_6}\ket{\alpha}_{\phi_7}\ket{5\alpha}_{\phi_8}$, with the intensity distribution shown in Fig.~\ref{fig:optical_switch} (b). After propagating through the MPLC system configured to perform the eight-mode identity transformation and an eight-mode permutation transformation, the output field's intensity distribution are recorded with the EMCCD camera and shown in Fig.~\ref{fig:optical_switch} (b). The performance is quantified with normalized power matrices.

Fig.~\ref{fig:subspace_BS} shows the results demonstrating the MPLC performing subspace two-mode beam splitter in a 4-dimensional (panel (a)) and 8-dimensional (panel (b)) Hilbert space. The subspace beam splitters can be formally defined with the target unitaries shown in Fig.~\ref{fig:subspace_BS}. It performs a $50:50$ beam splitter on two of the pixel modes while keeping the rest of the pixel modes intact. For the four-mode subspace beam splitter, we prepare an input state, $\ket{\alpha}_{\phi_1}\ket{\alpha}_{\phi_2}\ket{\alpha}_{\phi_3}\ket{\alpha}_{\phi_4}$, whose intensity distribution is recorded with a CMOS camera. After propagating through the MPLC system configured to perform the target unitary, the output field intensity distribution is recorded with the EMCCD camera. Similarly, for the eight-mode subspace beam splitter transformation, we prepare an input state $\ket{\alpha}_{\phi_1}\ket{\alpha}_{\phi_2}\ket{\alpha}_{\phi_3}\ket{\alpha}_{\phi_4}\ket{\alpha}_{\phi_5}\ket{\alpha}_{\phi_6}\ket{\alpha}_{\phi_7}\ket{\alpha}_{\phi_8}$, with the intensity distribution shown in Fig.~\ref{fig:subspace_BS} (b). After the transformation, the output field's intensity distribution is recorded with the EMCCD camera. We quantify the performance with the normalized power matrices.

\section{Conclusion}~\label{sec:conclusions}

A central conclusion of this work is that programmable multi-plane light conversion can operate natively as a linear-optical interferometer on spatially localized pixel modes, with the required plane count for high-fidelity compilation scaling approximately linearly with the mode count of the desired unitary to be realized on the pixel mode array. In experiments, we demonstrate programmable interferometers operating on up to sixteen modes that were inspired by various applications in classical and quantum information processing. 

In future work, it would be instructive to do actual demonstrations of some of the quantum photonic use cases of a pixel-mode diffractive unitary, such as discrete and continuous-variable cluster-state generation~\cite{Pant2019} via boosted-BSM based fusions~\cite{Ewert2014}: perhaps also combining it with programmable multi-time-bin linear optical transformations~\cite{Cui2025,Basani2026}, Gaussian Boson Sampling~\cite{Hamilton2017}, Boson Sampling~\cite{Aaronson2011}, and more. In addition to above, this capability could also be used for free-space scalable optical control and manipulation of trapped ion and tweezer-trapped atoms\cite{manetsch2025tweezer,holman2026trapping}. 

\section*{Author Contributions}
MS implemented the wavefront matching algorithm in Python, and performed all the numerical simulations of the unitaries in this paper, including the development of the plane-count scaling results. IO built the MPLC at UMD---while upgrading a previous reconfigurable MPLC experiment led by MRG at UArizona---using a MATLAB code that MRG wrote for Ref.~\cite{Ozer2022} which IO repurposed to work for pixel-mode unitary transformation; and made the new MPLC design to be compatible with an external light source. IO, MS and WH collaborated to harmonize geometrical assumptions in MS's Python code to make it work with the experimental setup. MRG conceptualized interpreting the pixel-mode unitary as a spatial mode sorter, and did initial simulations for the two-mode splitter. WH conceptualized the design of the complex-amplitude modulation SLM-based reconfigurable coherent scene generator, and built it in collaboration with IO. WH also advised MS on numerical computations pertaining to form-factor evaluations. CC provided guidance on the experiment design, supported debugging on the simulation and experiment, and revised the manuscript. SG formulated the problem and its use cases, provided overall guidance to the research, and assembled the article with inputs from MS and IO.

\begin{acknowledgements}
The authors acknowledge Mark Meisner of RTX for useful discussions, and also for funding support under a RTX-UMD master research agreement.
\end{acknowledgements}

\section*{Code and Availability}

The code that supports the findings of this study is openly available~\cite{sureka2026mplcgithub}. The data are available from the authors upon reasonable request.

\bibliography{MPLCunitary}

\appendix

\section{Wavefront matching for interferometric realization of a multimode pixel-mode unitary}~\label{app:wavefrontmatching_unitary}
Our MPLC device consists of a series of $K$ spatial light modulator (SLM) planes, each separated by a distance $w$ from the next one. The SLM planes are P $\times$ P pixelated masks with pixel size, $\Delta \times \Delta$. Each (square-shaped) pixel has a center $(u_i,v_j)$ where $i$ and $j$ range from $1$ to $P$. We assume unity fill factor. Therefore, the linear dimension of the mask, $H = P \times \Delta$. In our problem context, the input field is constrained to be a superposition of $M$ pixel (circ-function) modes centered at $(x_m,y_m)$, $1 \le m \le M$, with mode-field diameters $\sigma$. The $M$ pixel modes at the input plane ($z=0$) are given by $\phi_m(x,y) \equiv \phi_0(x-x_m,y-y_m)$, where the centered-circ function,
\begin{equation}
\label{mode func}
\phi_0(x, y) = 
\begin{cases}
\frac{1}{\sqrt{A}}, & \text{if } x^2 + y^2 \leq \frac{\sigma^2}{4} \\
0, & \text{otherwise}
\end{cases},
\end{equation}
where \( A = {\pi \sigma^2}/{4} \) is the area of each pixel mode. We denote the $m$-th pixel-mode region in the $(x,y)$ plane, as ${\cal P}^{\rm (in)}_m$. At the MPLC output plane, the $M$ pixel modes are arranged on an identical Cartesian grid in the $(x', y')$ plane at $z=L$, i.e., $\psi_m(x^\prime,y^\prime) \equiv \phi_0(x^\prime-x_m,y^\prime-y_m)$ (see Fig.~\ref{fig:bulge}). We denote the $m$-th pixel-mode region in the $(x',y')$ plane, as ${\cal P}^{\rm (out)}_m$, $1 \le m \le M$. We express the $M$-mode unitary $U$ we wish to realize, as:
\begin{equation}
{\hat {\boldsymbol a}_\mathrm{out}} = U {\hat {\boldsymbol a}_\mathrm{in}},
\end{equation}
where ${\hat {\boldsymbol a}_\mathrm{out}}$ and ${\hat {\boldsymbol a}_\mathrm{in}}$ are column vectors comprising the modal annihilation operators ${\hat a}_{\mathrm{in},m}$ and ${\hat a}_{\mathrm{out},m}$, $1 \le m \le M$, that we assign to the input-plane and output-plane pixel mode sets, respectively. 

\subsection{Pixel mode unitary as a spatial mode sorter}~\label{app:unitary_as_modesorter}
Next, we will argue how we can re-interpret the MPLC-based unitary transformation as a spatial-mode sorter. If we populate the input modes in coherent states $|\alpha_m\rangle$, $1 \le m \le M$, the output modes will be excited in coherent states $|\beta_m\rangle$, $1 \le m \le M$, where the column-vector coherent-state complex amplitudes at the output and input are related via: ${\boldsymbol \beta} = U {\boldsymbol \alpha}$. Let us call the total mean photon number across the pixel modes, $N = \sum_{m=1}^M |\alpha_m|^2 = \sum_{m=1}^M |\beta_m|^2$. Let us denote the $(i,j)$-th element of $U$ as $u_{ij}$ with $i$ and $j$ denoting the row and column indices respectively. The $(i,j)$-th element of $U^\dagger$ is therefore $u_{ji}^*$. Therefore, if we want the output to be ${\boldsymbol \beta} = [0, \ldots, 0, \gamma, 0, \ldots, 0]^{\rm T}$, with only the $n$-th term non-zero, the input coherent state amplitudes must be given by: ${\boldsymbol \alpha} = U^\dagger {\boldsymbol \beta} = [u_{n1}^*, u_{n2}^*, \ldots, u_{nM}^*]^{\rm T}\gamma$. In other words, if we were to excite the input facet of our pixel-mode MPLC unitary in a coherent state $|\gamma\rangle$ of the spatial mode:
\begin{equation}
\zeta_n(x,y) \equiv \sum_{m=1}^M u_{nm}^* \phi_m(x,y),
\end{equation}
the field at the output facet of the MPLC will be excited in the coherent state $|\gamma\rangle$ of the mode $\psi_n(x^\prime,y^\prime)$, and the rest of the output pixel modes, $\psi_k(x^\prime,y^\prime), k \ne n$, will be in vacuum. Let us consider the mode-overlap integral:
\begin{eqnarray}
c_{kl} &\equiv& \int \zeta_k(x,y) \zeta_l^*(x,y)dxdy, (k,l) \in (1, \ldots, M) \nonumber \\
&=& \int \sum_{m=1}^M u_{km}^* \phi_m(x,y) \sum_{n=1}^M u_{ln} \phi_n^*(x,y) dxdy \nonumber \\
&=& \sum_{m=1}^M \sum_{n=1}^M u_{km}^* u_{ln}\int \phi_m(x,y) \phi_n^*(x,y) dxdy \nonumber \\
&=& \sum_{m=1}^M \sum_{n=1}^M u_{km}^* u_{ln} \delta_{mn} \nonumber \\
&=& \sum_{m=1}^M u_{km}^* u_{lm} \nonumber \\
&=& \sum_{m=1}^M (U^\dagger)_{\{mk\} } U_{\{lm\}} \nonumber \\
&=& ( UU^\dagger)_{\{kl\}} \nonumber \\
&=& ({\mathbb I}_M)_{\{kl\}} \nonumber \\
&=& \delta_{kl},
\end{eqnarray}
which uses the orthogonality of the input pixel modes (which naturally stems from them being spatially non-overlapping) to prove that the spatial modes $\zeta_m(x,y)$, $1 \le m \le M$ are mutually orthogonal.

This lets us re-interpret the MPLC based pixel-mode unitary transformation as the more traditional view of the MPLC being used as a spatial-mode sorter that spatially separates the mutually orthogonal spatial modes $\zeta_m(x,y)$, $1 \le m \le M$, into the spatially-separated (and hence orthogonal) output-plane pixel modes $\psi_m(x^\prime,y^\prime)$, $1 \le m \le M$, respectively. With this viewpoint, we can now borrow techniques from iterative wavefront-matching techniques from the literature to compile our MPLC's SLM phases~\cite{fontaine2019}, which we describe next.

\subsection{Iterative wavefront matching algorithm}~\label{app:wavefrontmatching}

Our MPLC design in Fig.~\ref{fig:bulge} comprises $K$ spatial-light modulator (SLM) planes. We take the first plane $(k=1)$ to be coincident with the object plane $(x,y)$, but we allow for a tunable propagation length between the last plane $(k=K)$ and the focal plane camera. The forward propagation in Fig.~\ref{fig:bulge} is seen to go from (left to right) $z=0$ (object plane) to $z=w$ (plane 1), $z=2w$ (plane 2), etc., to $z=Kw$ (plane K), and finally on to $z = Kw + w_L \equiv L$, where $w_L$ is the final propagation segment we allow from the $K$-th plane to the camera. 

The right-propagating field to the {\em left} of the $k$-th plane when one of the input-plane modes $\zeta_m(x,y)$ is transmitted, $1 \le m \le M$, during the $j$-th iteration of the wavefront matching (WFM) algorithm, is denoted $A_{km}^{(j)}(x,y)$. The SLM phase for the $k$-th plane for the $j$-th iteration is taken to be $\theta_k^{(j)}(x,y)$. Therefore, the right-propagating field to the {\em right} of the $k$-th plane for mode $m$, $1 \le m \le M$, for the $j$-th iteration of the algorithm is $A_{km}^{(j)}(x,y)e^{-i \theta_k^{(j)}(x,y)}$. In tandem, we left-propagate the desired output-plane (pixel) modes $\psi_m(x^\prime,y^\prime)$ backwards through the MPLC planes. For this back-propagation step, we name the field on the right side of plane $k$ during iteration $j$ as $B_{km}^{(j)}(x,y)$.

The intuition behind the WFM algorithm\cite{fontaine2019} is to pick SLM phases at each of the planes, $\theta_k^{(j)}(x,y)$, such that they `work for' all $m \in \left\{1,\ldots,M\right\}$, viz., in mapping the modes $\zeta_m(x,y)$ at the input to modes $\psi_m(x^\prime,y^\prime)$ at the output, simultaneously. In other words, after convergence ($j$ large enough), we wish $A_{km}^{(j)}(x,y)e^{-i \theta_k^{(j)}(x,y)} = B_{km}^{(j)}(x,y)$, $\forall m, k$; or $AB^* = (BB^*) e^{i \theta}$ (dropping all indices), implying $\theta = {\text{arg}}(AB^*)$. 

Armed with the above intuition, let us start by initializing the iteration counter $j=0$, and initialize all the SLM phases to $\theta_k^{(0)}(x,y) = 0$, such that $\exp[i\theta_k^{(0)}(x,y)]=1$, $\forall k \in \left\{1, \ldots, K\right\}$. Next, for a fixed total number of iterations, $0 \le j \le J$,
\begin{itemize}
\item We perform one step of {\em forward propagation} for all the $M$ modes $\zeta_m(x,y)$ at all the $K$ masks, i.e., initialize $A_{1,m}^{(0)}(x,y) = \zeta_m(x,y)$, and compute $A_{k+1,m}^{(j)}(x,y) = {\cal R}_w\big[A_{k,m}^{(j)}(x,y) e^{-i \theta_k^{(j)}(x,y)}\big]$, where 
\begin{align}
& \qquad {\cal R}_w\big[E(x,y)\big] \equiv \nonumber \\
& \qquad \frac{1}{i\lambda} \iint_{-\infty}^{\infty} E(x', y') \frac{w e^{i 2\pi r/\lambda}}{r^2}\left(1+\frac{i \lambda}{2\pi r}\right) dxdy
\end{align}
is the $\lambda$-center-wavelength Rayleigh-Sommerfeld diffractive propagation of a quasi-monochromatic field $E(x,y)$ over a distance $w$ in the direction of the Poynting's vector perpendicular to the $(x,y)$-plane, and $r = \sqrt{(x-x')^2 + (y-y')^2 + w^2}$ is the radial Euclidean distance between the points $(x',y',0)$ and $(x,y,w)$, the third coordinate being $z$.

\item We perform one step of {\em backward propagation} for all the $M$ output-plane pixel modes $\psi_m(x,y)$ at all the $K$ masks, i.e., initialize $B_{K,m}^{(0)}(x,y) = \psi_m(x,y)$ (here, we are assuming $w_L = 0$ for simplicity of exposition), and compute $B_{k-1,m}^{(j)}(x,y) = {\cal R}_{-w}\big[B_{k,m}^{(j)}(x,y) e^{i \theta_k^{(j)}(x,y)}\big]$. Note that in writing the above, we are abusing notation by using $(x,y)$ to denote the transverse coordinates at each of the $K$ planes.
\item Update phase masks at each of the $K$ planes, as:
\begin{equation}
\quad \theta_k^{(j+1)}(x,y) = {\text{arg}}\left(\sum_{m=1}^M A_{k,m}^{(j)}(x,y)B_{k,m}^{(j)*}(x,y)\right).
\end{equation}

\item Increment iteration counter $j \to j+1$, and repeat the above three steps, until $j = J$; and then stop.
\end{itemize}

\subsection{Wavefront matching (WFM) convergence}~\label{app:wavefrontmatchingconv}

\begin{figure}
    \centering
    \includegraphics[width=0.9\columnwidth]{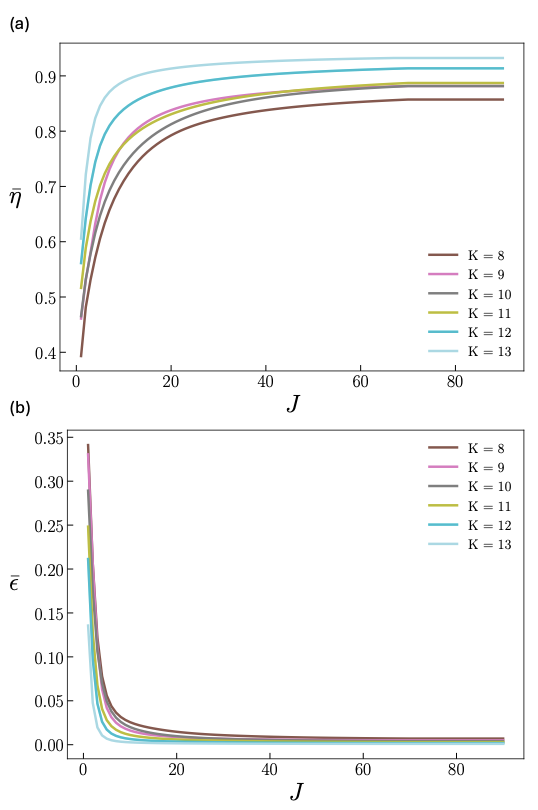}
    \caption{(a) Mean transmissivity $\bar{\eta}$ and (b) mean crosstalk ${\bar \epsilon}$, vs. iteration count $J$, for the $16$-mode Hadamard unitary, with plane count $K$ $= 8$ to $13$.}
    \label{fig:performance_iterations}
\end{figure}

To assess the performance of the wavefront matching algorithm (described in Appendix~\ref{app:wavefrontmatching}), we numerically evaluate $\bar{\eta}$ and $\bar{\epsilon}$ over the course of $J \le 90$ WFM iterations for various values of plane-count $K$. The transmissivity is seen to consistently improve with iteration count $J$, exhibiting faster convergence for larger values of $K$ due to increased degrees of freedom in the phase modulation (see Figure~\ref{fig:performance_iterations}(a)). Concurrently, the crosstalk decreases steadily with $J$ (see Figure~\ref{fig:performance_iterations}(b)), demonstrating enhanced spatial mode selectivity and reduced leakage. Together, these metrics validate the effectiveness of the wavefront matching process and inform the design trade-offs associated with selecting $K$. Transmissivity \((\bar{\eta})\) is seen to rise rapidly in the first \(30\) iterations and then increases slowly until saturating at \(\sim 65\) iterations. Crosstalk \((\bar{\epsilon})\) drops sharply and plateaus on the same timescale. Larger plane counts $K$ converge faster and reach higher \(\bar{\eta}\) with lower \(\bar{\epsilon}\), reflecting more degrees of freedom for wavefront shaping. For all of the numerical results in the main paper, we therefore used $J=60$ iterations to evaluate performance.

\section{Definitions of performance metrics: transmissivity and crosstalk}~\label{app:definitions}

We now define the average transmissivity \( \bar{\eta} \) and the average crosstalk \( \bar{\epsilon} \) that help us assess the performance of implementing a particular pixel-mode unitary. 

\subsection{Transmissivity}
Since the mode functions $\zeta_m(x,y)$ are unit-norm, the normalized input energy, $I^{\rm in}_m = \int\int |\zeta_m(x,y)|^2dxdy = 1$. We define the transmissivity of the $m$-th transmitted mode as:
\begin{equation}
\eta_m = \frac{I^{\rm out}_m}{I^{\rm in}_m} \le 1,
\end{equation}
where
\begin{equation}
I^{\rm out}_m = \sum_{l=1}^M\int\int_{(x,y) \in {\cal P}_l^{\rm out}} |A_{K,m}^{(J)}(x,y)|^2 dxdy \le 1
\end{equation}
is the normalized output energy: the fraction of the input energy in the $\zeta_m$ mode (which is by definition spread across all $M$ input-plane pixel mode regions ${\cal P}^{\rm (in)}_m$) that enters across {\em all} of the $M$ output-plane pixel mode regions ${\cal P}^{\rm (out)}_l$. The average transmissivity is given by:
\begin{equation}
{\bar \eta} = \frac1M\sum_{m=1}^M \eta_m.
\end{equation}

\subsection{Crosstalk}
The crosstalk associated with a given input mode $\zeta_m(x,y)$ is defined as the fraction of the total light intensity that couples into unintended output modes, relative to the total intensity distributed across all output modes. The average crosstalk, denoted $\bar{\epsilon}$ is obtained by averaging the crosstalk over all $M$ input modes. Hence:
\begin{equation}
\bar{\epsilon} = \frac{1}{{M}} \sum_{m=1}^{{M}} \epsilon_m,
\end{equation}
where
\begin{equation}
1-\epsilon_m = \frac{1}{I_m^{\rm out}}{\int\int_{(x,y) \in {\cal P}_m^{\rm out}} |A_{K,m}^{(J)}(x,y)|^2 dxdy}.
\end{equation}

\section{Performance evaluation of various use-case inspired pixel-mode unitaries}\label{app:examples}

\begin{figure}
    \centering
    \includegraphics[width=\columnwidth]{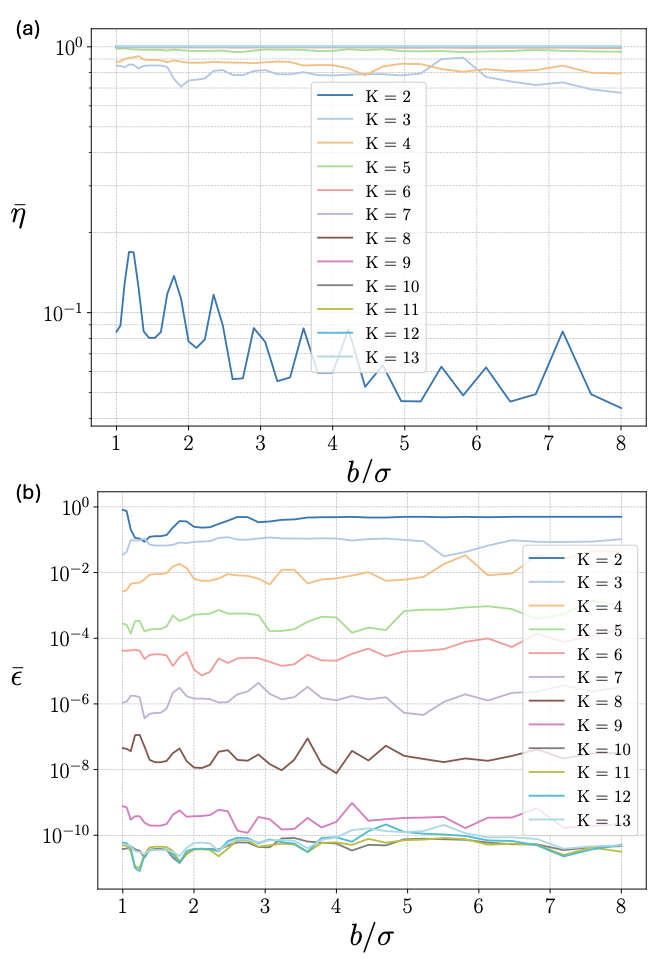}
    \caption{(a) Mean transmissivity $\bar{\eta}$ and (b) mean crosstalk ${\bar \epsilon}$, plotted as a function of MFD-normalized inter-pixel-mode spacing $b/\sigma$, for the alternative 50-50 beamsplitter unitary in Eq.~\eqref{eq:altbeamsplitter}.}
    \label{fig:performance_altbeamsplitter}
\end{figure}

\begin{figure}
    \centering
    \includegraphics[width=\columnwidth]{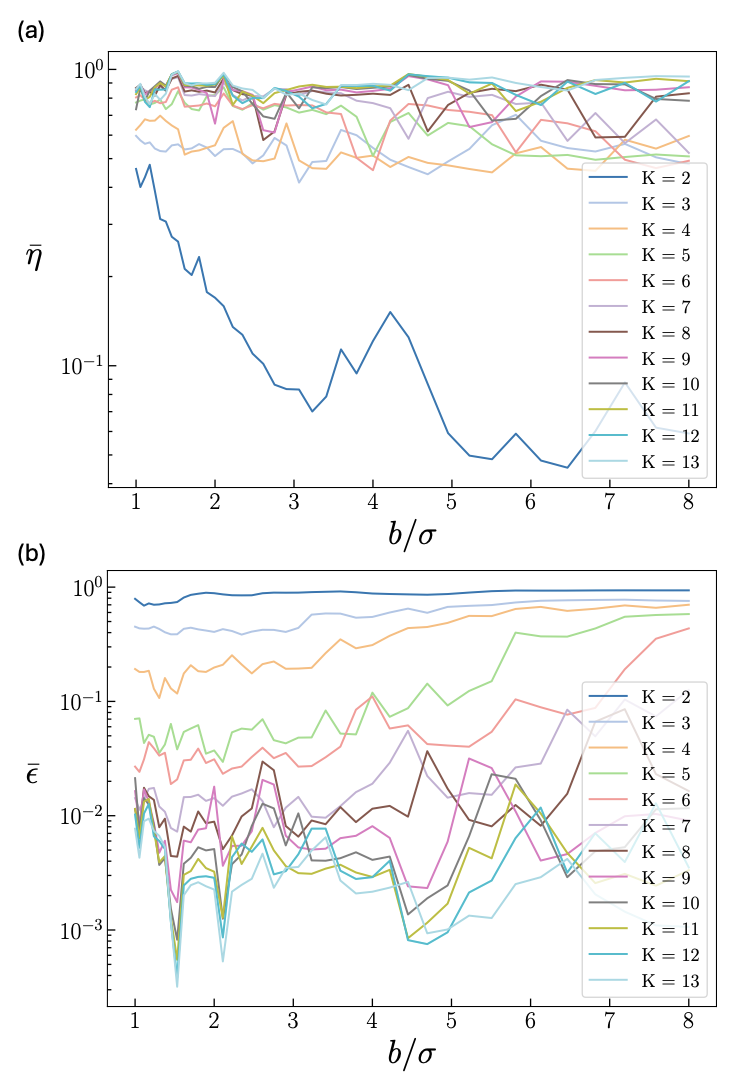}
    \caption{Numerical performance evaluationf of transmissivity and crosstalk for the $16$-mode real Hadamard unitary.}
    \label{fig:performance_16_mode_hadamard}
\end{figure}

\begin{figure*}
    \centering
    \includegraphics[width=\textwidth]{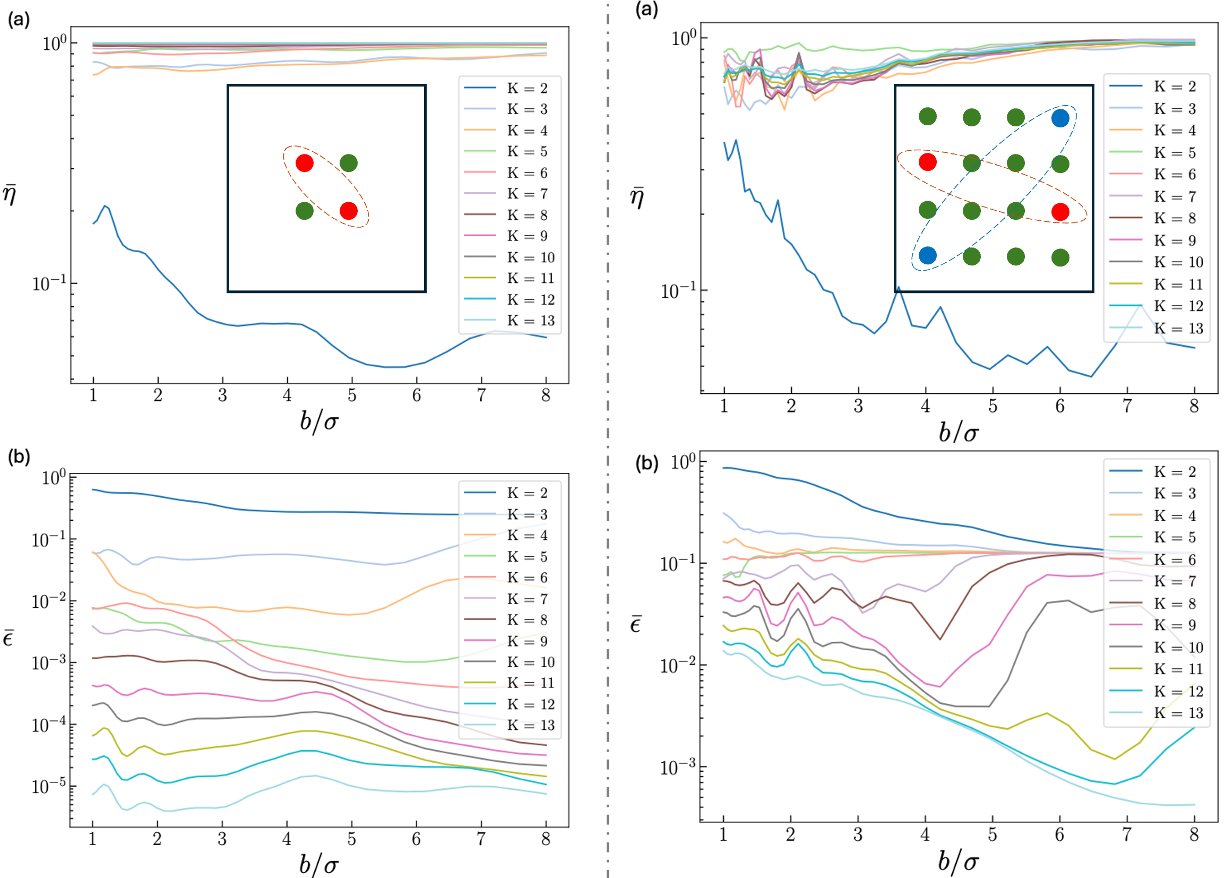}
    \caption{Numerical performance evaluations for two examples of partial unitaries: (left panel) a 2-mode beamsplitter among two of the four modes of a $4$-pixel-mode grid, and (right panel) two simultaneous 2-mode beamsplitters given by Eq.~\eqref{eq:beamsplitter} among two pairs of modes of a $16$-pixel-mode grid.}
    \label{fig:partialbeamsplitters}
\end{figure*}

In this Appendix, we plot the average transmissivity and crosstalk, as a function of $b/\sigma$, for the following use-case inspired pixel-mode unitaries:

\begin{enumerate}
\item An alternative 2-mode beamsplitter unitary shown in Eq.~\eqref{eq:altbeamsplitter} (see Fig.~\ref{fig:performance_altbeamsplitter}), 
\begin{equation}
U = \frac{1}{\sqrt{2}}\left(\begin{array}{cc}
1 & i \\
i & 1
\end{array}\right).
\label{eq:altbeamsplitter}
\end{equation}

\item An $8$-mode linear optical circuit which, with $4$ ancilla photons, can attain in principle a dual-rail photonic-qubit-basis Bell State Measurement (BSM) whose success probability is $3/4$~\cite{Ewert2014} (see Fig.~\ref{fig:performance_boostedBSM8}), and \item A $16$-mode linear optical circuit which, with $12$ ancilla photons, can attain in principle a dual-rail photonic-qubit-basis Bell State Measurement (BSM) whose success probability is $25/32$~\cite{Ewert2014} (see Fig.~\ref{fig:performance_boostedBSM16}).

\item The $16$-mode real Hadamard unitary (see Fig.~\ref{fig:performance_16_mode_hadamard}).

\item Two instances of partial beamsplitters: (a) One 2-mode beamsplitter acting on a diagonally-opposite mode pair in a $2 \times 2$ 4-pixel-mode grid, and (b) two simultaneous 2-mode beamsplitters acting on two pixel-mode pairs in a $4 \times 4$ 16-pixel-mode grid (see Fig.~\ref{fig:partialbeamsplitters}). 
\end{enumerate}

\begin{figure}
    \centering
    \includegraphics[width=0.8\columnwidth]{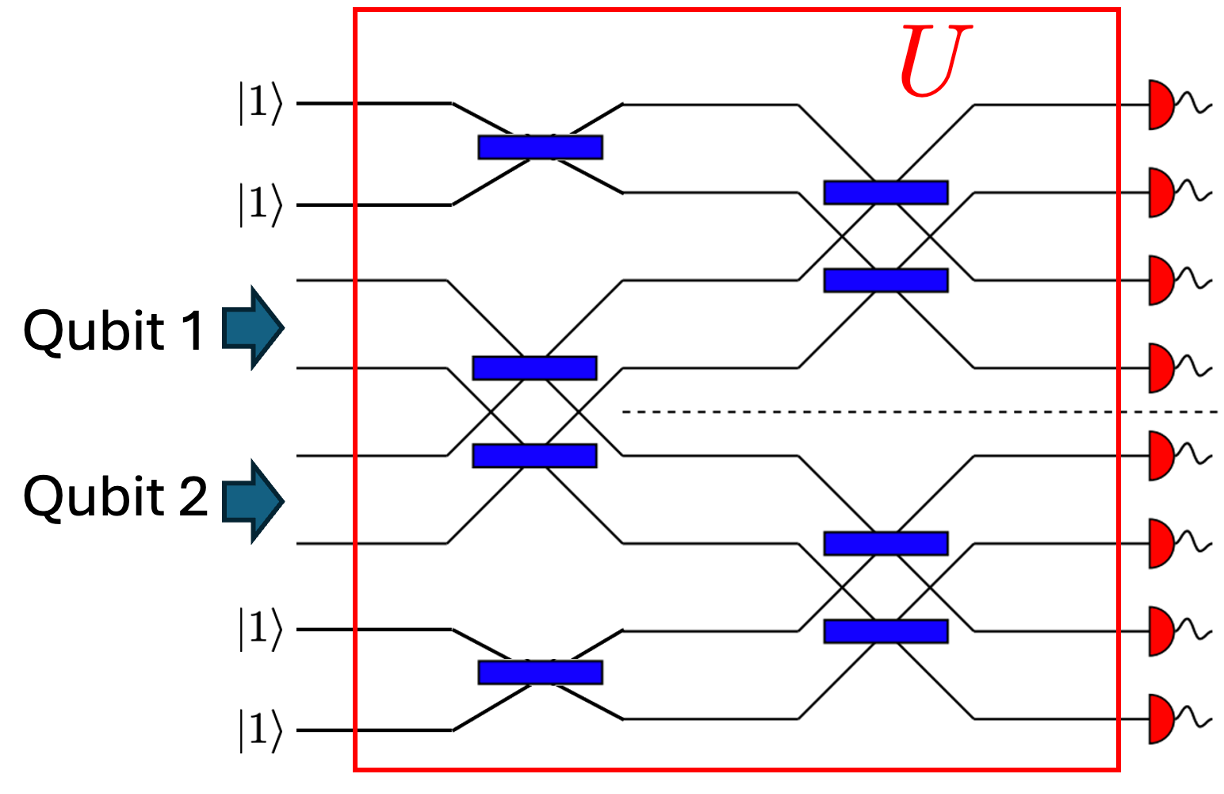}
    \caption{
     A linear optical circuit implementing an 8-mode unitary that realizes, under ideal circumstances, a 3/4-success-probability Bell-state measurement (BSM) on dual-rail photonic Bell states. Each 2-mode beamsplitter is given by the unitary in  Eq.~\eqref{eq:altbeamsplitter}.}
    \label{fig:boostedBSM8}
\end{figure}

\begin{figure}
    \centering
    \includegraphics[width=\columnwidth]{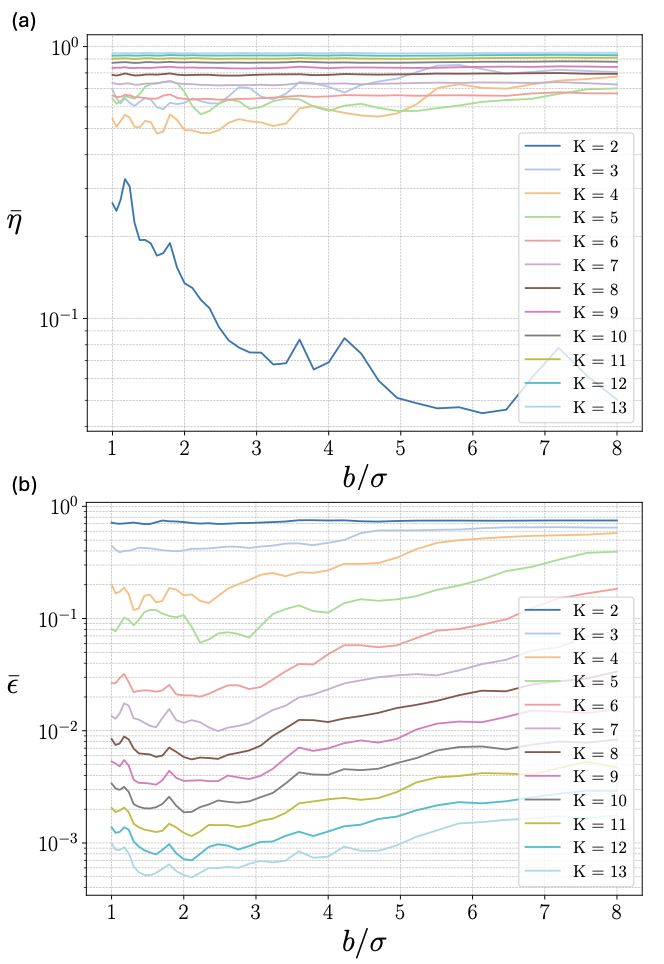}
    \caption{(a) Mean transmissivity $\bar{\eta}$ and (b) mean crosstalk ${\bar \epsilon}$, plotted as a function of MFD-normalized inter-pixel-mode spacing $b/\sigma$, for the $8$-mode unitary for a $3/4$-success probability Boosted BSM, shown in Fig.~\ref{fig:boostedBSM8}.}
    \label{fig:performance_boostedBSM8}
\end{figure}
\clearpage
\begin{figure}
    \centering
    \includegraphics[width=\columnwidth]{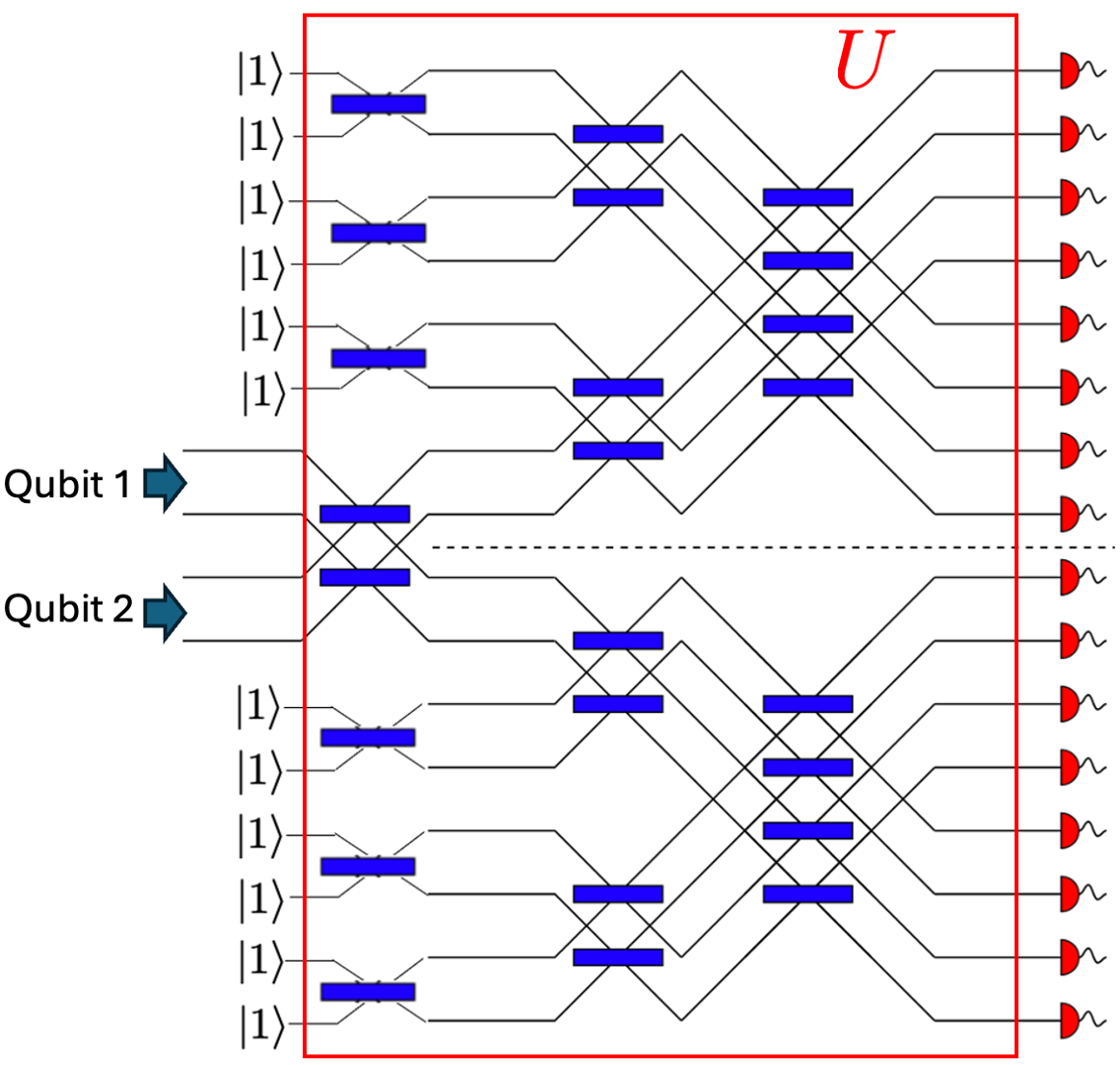}
    \caption{A linear optical circuit implementing a $16$-mode unitary that realizes, under ideal circumstances, a $25/32$-success-probability Bell-State Measurement (BSM) on dual-rail photonic Bell states. Each $2$-mode beam splitter is given by the unitary in Eq.~\eqref{eq:altbeamsplitter}.}
    \label{fig:boostedBSM16}
\end{figure}
\newpage
\begin{figure}
    \centering
    \includegraphics[width=\columnwidth]{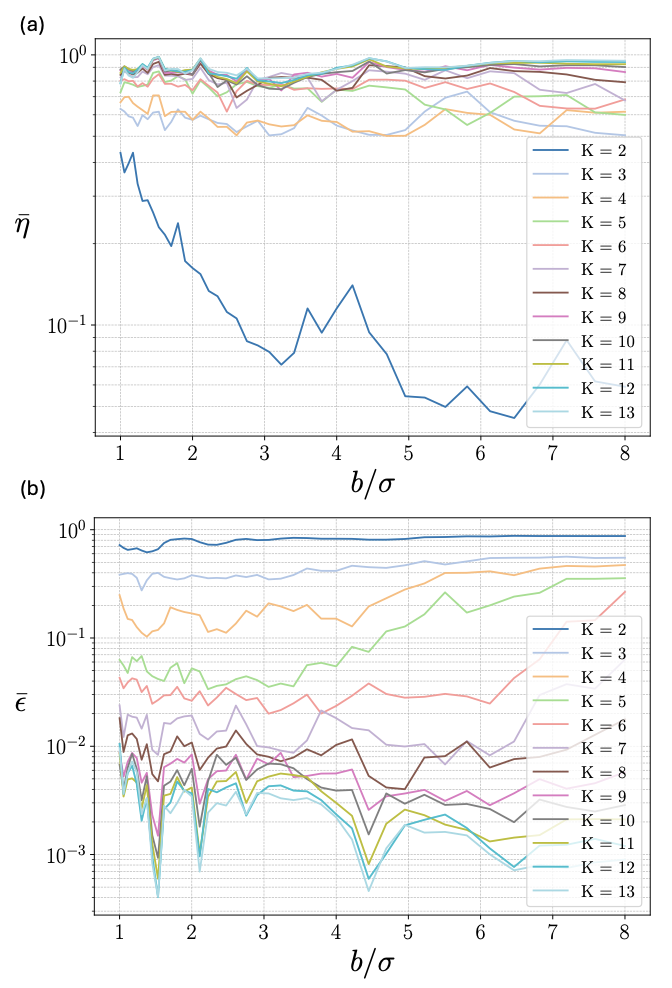}
    \caption{(a) Mean transmissivity $\bar{\eta}$ and (b) mean crosstalk ${\bar \epsilon}$, plotted as a function of MFD-normalized inter-pixel-mode spacing $b/\sigma$, for the $16$-mode unitary for a $25/32$-success probability Boosted BSM, shown in Fig.~\ref{fig:boostedBSM16}.}
    \label{fig:performance_boostedBSM16}
\end{figure}
\clearpage
\clearpage

\end{document}